# Versatile Non-diffracting Perfect Vortex Beam

Wenxiang Yan,[1] Yuan Gao,[1] Zheng Yuan,[1] Zhe Weng,[1] Zhi-Cheng Ren,[1] Xi-Lin Wang,[1] Jianping Ding,[1,2,3,*] and Hui-Tian Wang[1]

[1]Nanjing University, National Laboratory of Solid Microstructures and School of Physics, Nanjing, China
[2]Nanjing University, Collaborative Innovation Center of Advanced Microstructures, Nanjing, China
[3]Nanjing University, Collaborative Innovation Center of Solid-State Lighting and Energy-Saving Electronics, Nanjing, China

*Corresponding author: jpding@nju.edu.cn

The rapid scale broadening and divergence increasing of vortex beams (VBs) with orbital angular momentum (OAM), e.g., Laguerre-Gaussian beams, severely impede the wide applications of VBs ranging from optical manipulation to high-dimensional quantum information communications, which call for VBs to have the same transverse scale and divergence for distinct OAM or even the small vortex ring for large OAM. Non-diffracting beams, on the other hand, that are capable of overcoming diffraction without divergence, are very evocative and indeed appealing in numerous applications including atom optics and medical imaging. Here, we propose theoretically and demonstrate experimentally a brand new type of VB having OAM-independent radii meanwhile holding propagation-invariant without divergence as well as self-healing properties, named non-diffracting perfect vortex beam (NDPVB). We work out a versatile toolkit based on Fourier-space analysis to multidimensionally customize NDPVBs at will so that it is of propagating intensity and phase controllability with intriguing customizable behaviors of self-accelerating, self-similar, and self-rotating. This goes beyond tailoring the transverse plane to the higher-dimensional propagating characteristics in structured light beams. A deeper insight into the internal flow revealed and confirmed that the multidimensional customization of NDPVBs is dominated by inducing corresponding multidimensional internal flow, facilitating our understanding of how our design scheme of propagating properties manipulates the internal flows, unveiling the nature of structure formation and behavior transformation of structured light beams. The arbitrarily customized NDPVBs offer promising alternatives for the classic VBs in numerous application scenarios and the versatile toolbox based on the Fourier space can enrich the toolkit for beams' shaping and give new insight into the evolution of the light field.

## Remove the inherent limitation of VBs

Vortices are ubiquitous phenomena in nature, from quantum vortices in liquid nitrogen to typhoon vortices and even to spiral galaxies in the Milky Way, manifesting themselves not only in macroscopic matter but also in structured electromagnetic fields. VBs, optical vortex fields propagating in the paraxial regime and carrying OAM proposed by Allen et al.[1] in 1992, are characterized by a spirally increasing (decreasing) phase in the form of $\exp(im\varphi)$, where $m$ is an integer referred to as a topological charge (TC), $\varphi$ is the azimuthal angle. The OAM carried by VBs is equivalent to $m\hbar$ per photon ($\hbar$ is the Dirac constant) and can be much greater than the spin angular momentum (SAM) of $\hbar$ carried by a photon. Since the values of OAM are theoretically unbounded in VBs, applications of the VBs are multitudinous in high-dimensional classical and quantum information communications[2], micro-particle manipulation[3], optical measurements[4], optical imaging[5] and processing[6], and so on.

Carrying the aforementioned spiral phase, VBs manifest as a ring shape with a phase singularity at its center[7] and the radius of VBs is always positively correlated with OAM (e.g., Laguerre-Gaussian beams and Bessel beams shown in Fig. 1a). Namely, the radii of VBs carrying greater OAM are certainly much larger than those carrying smaller OAM at the same conditions. This inherent 'unperfectness' property (i.e., the OAM-dependent size) will restrict the application in scenarios that use VBs with distinct OAMs simultaneously, like fiber optic data transmission, spatial OAM mode (de)multiplexing communication, and particle manipulation. Less mature, but receiving intense interest, is the attempt to lift this restriction. However, conventional perfect optical vortex fields proposed until now are merely restricted to a single cross-sectional plane distribution and cannot maintain perfection when propagating[8,9]. Leaving away from this certain plane, the divergence of conventional perfect optical vortex fields increases rapidly as the OAM enlarges, and thus we refer to these as two-dimensional (2D) perfect vortex fields. To the best of our knowledge, there is no report on the three-dimensional (3D) perfect vortex field, or perfect vortex beam (PVB) which can continue to be perfect even after propagation. To address this appealing issue, we present a brand new type of VB to completely remove this inherent 'unperfectness' restriction of VBs, as illustrated in Fig. 1b. By solving the distributions of high-order Bessel beams (HOBBs), we work out an analytical expression for the vortex radius and discover that this vortex radius is not only positively related to the OAM (TC) but also inversely proportional to the radial wavevector component. Using the derived analytical expression, we generate the new VB whose vortex radius can be accurately designed arbitrarily regardless of the OAM carried. This VB is propagation-invariant without divergence and can self-heal after suffering perturbed or impaired, thereby maintaining 'perfection' while propagating, and thus is named a non-diffracting perfect vortex beam (NDPVB). The theoretical derivation is as follows.

Light diffraction is one of the universal phenomena and is intricately linked to the wave nature of light. In 1987, Durnin[10] established a class of solutions known as non-diffracting Bessel beams, which have electric field amplitudes proportional to Bessel functions and intensity patterns that are propagation-invariant. The field distribution of an $m$th-order Bessel beam in cylindrical coordinates $\boldsymbol{r}=(r,\varphi,z)$ is given by

$$U(r,\varphi,z) = J_m(k_{r0}r)e^{im\varphi}e^{ik_{z0}z}, \qquad (1)$$

where $J_m$ is the $m$th-order Bessel function, and $k_{z0}$, $k_{r0}$ are the longitudinal and radial components of the wavevector, such that $k_0 = 2\pi/\lambda = \sqrt{k_{r0}^2 + k_{z0}^2}$ and $\lambda$ is the wavelength (assuming 532 nm in this paper). The intensity pattern of non-diffracting HOBB is independent of $z$ in the sense that $I(r,\varphi,z) = I(r,\varphi)$. As a type of conventional VB, the HOBBs whose radii are also positively correlated with OAM are subject to this common constraint. By analyzing equation (1), we get the vortex radius expressed by

$$r_m = \left[C_1\left(2|m|\sqrt{1+1/|m|}\right) + C_2\right]/k_{r0}, \qquad (2)$$

where $r_m$ is the vortex radius of $m$th-order Bessel beam, and $C_1$ and $C_2$ are the correction coefficients ($C_1$=0.5207, $C_2$=0.7730 in this article, and the full derivation of equation (2) can be found in Supplementary theory 1). Notably, the vortex radius of HOBB is not only positively correlated with the TC $m$ (as expected from the common constraint), but it is also proportional to $1/k_{r0}$. By consciously tuning radial wavevector $k_{rm}$ for different TC $m$, we can remove this constraint, enabling us to precisely design vortex radii needed at will regardless of the TC. The following Perfect Equation (PE) is the formula for calculating $k_{rm}$ for a given radius $r_m$ and order $m$:

$$k_{rm} = \left[C_1\left(2|m|\sqrt{1+1/|m|}\right) + C_2\right]/r_m. \qquad (3)$$

We set $r_m$=$r_0$ to demonstrate PE's ability to generate the desired NDPVB with an TC-independent radius while inheriting the non-diffracting and self-healing properties of Bessel beams. The distribution of NDPVB can be represented as

$$U_{NDPVB}(r,\varphi,z) = C_m^{Amp} J_m(k_{rm}r)e^{im\varphi}e^{ik_{zm}z} \qquad (4)$$

where $k_{rm} = \left[C_1\left(2|m|\sqrt{1+1/|m|}\right) + C_2\right]/r_0$, $k_{zm} = \sqrt{k_0^2 - k_{rm}^2}$ and $C_m^{Amp} = 1/\max(J_m) = \sqrt[3]{3.2823|m| + 4.0849}$ is

the normalized amplitude coefficient (the full derivation can be found in Supplementary theory 2).

The common constraint of conventional VBs is that when the TC $m$ increases, the vortex radius $r_m$ enlarges, as shown in Fig. 1a: the first row, HOBBs with different orders, is calculated from equation (1) with $k_{r0}=0.0047k_0$ and $m$=12, 17, 22, 26. The corresponding vortex radii calculated from equation (2) are 250.0 um, 334.5 um, 439.0 um, and 514.6 um, agreeing well with the numerical simulation results of HOBBs; The second row, LGBs with different orders $m$=12, 17, 22 and 26 but the same radial order 3 and beam waist 146.5um, have the corresponding vortex radii of 250 um, 312.5 um, 367.6 um, 406 um (the basic theory of LGB can be found in Supplementary theory 3). In stark contrast to Fig. 1a, the NDPVBs shown in Fig. 1b generated from equation (5) with $r_0$=250um and $m$=12,17,22,26 have the same radii $r_0$ with changing orders solving this common constraint of the conventional vortex beams. Corresponding experimental results can be found in Supplementary Experimental Demonstration 1 and the experimental setup can be found in Supplementary Experimental Method. The ability to arbitrarily design vortex beams' radius regardless of the TC to NDPVB without divergence will open up more future opportunities for this dynamic field of optical vortices, fostering both fundamental advancements and applications. For example, the next generation of high-capacity OAM optical communication will benefit a lot from NDPVB and may pave the way for the integration and miniaturization of related communication devices. Besides, NDPVBs inherit the self-healing property of Bessel beams, in the sense that they tend to reconstitute or reform themselves even when they have been severely perturbed or impaired, illustrated in Supplementary Experimental Demonstration 2, which make themselves so robust that will fuel the applications under disturbed turbid circumstances like atmospheric turbulence, seawater, and biological tissues.

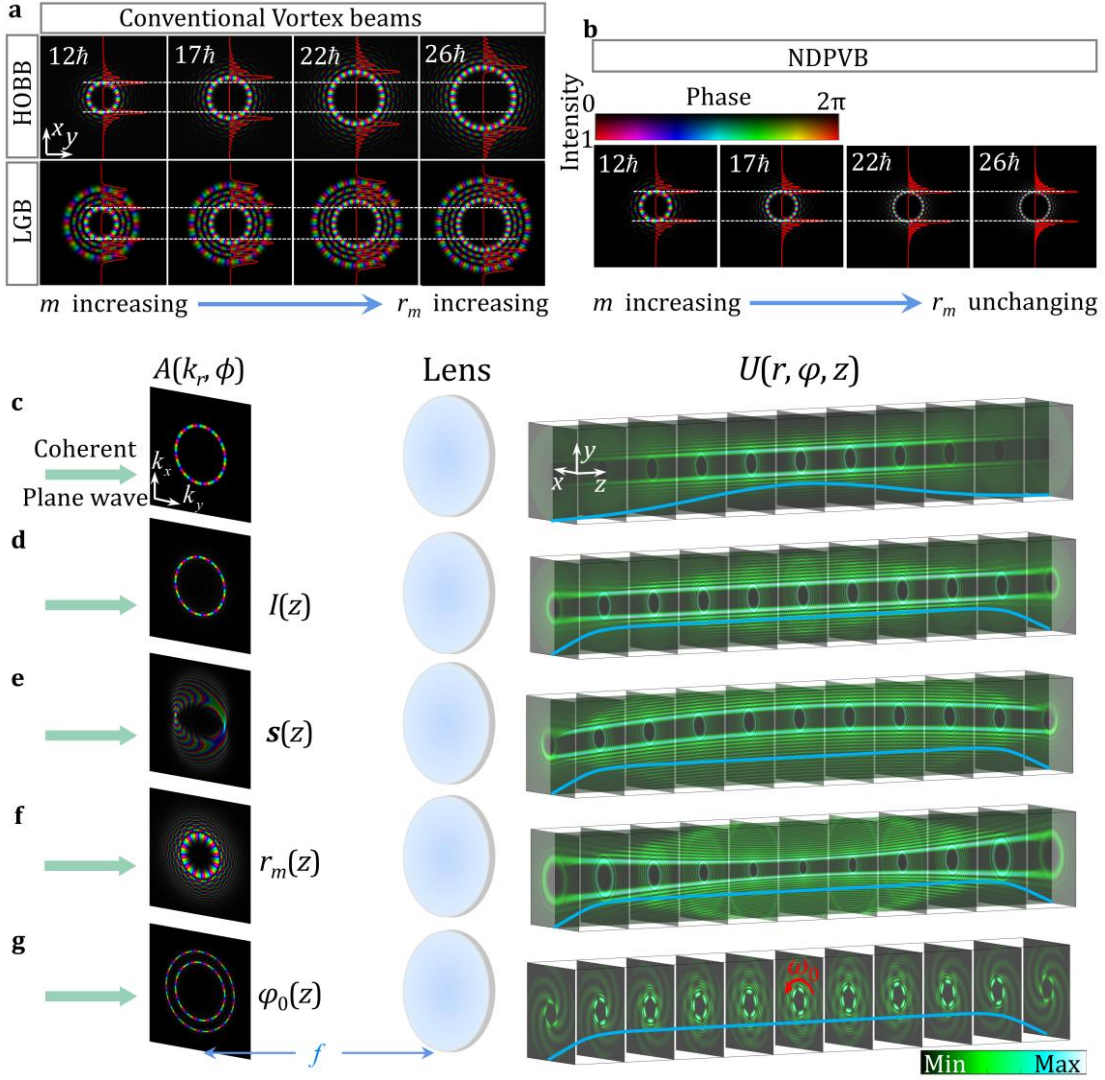

Fig.1 Schematic of NDPVBs and their multidimensional customization. The complex field distribution of **a** conventional VBs (LGBs and HOBBs) and **b** NDPVBs carrying OAM $12\hbar$, $17\hbar$, $22\hbar$, and $26\hbar$, respectively. The red curves represent intensity profiles along the x-axis and the horizontal white-dashed lines serve as a reference for the radius of the vortex beams carrying OAM $12\hbar$ in **a** and **b**. **c-g** Setting the angular spectrums $A(k_x, k_y)$ in the front focal plane of the converging lens with the focal length $f$ customizes NDPVBs in the focal volume $U(r, \varphi, z)$: **c** the quasi-NDPVB (i.e. Bessel Gaussian beams) with $f$ = 200 mm, $m$ = 12, the same hereinafter; **d** the customized NDPVB with uniform propagating intensity $I(z)$=rect($z/2f$); **e** the self-accelerating NDPVB with the trajectory $s(z)$ = 1.8×10$^{-4}$ [0, 1-$z^2/f^2$], **f** the self-similar NDPVB with changing radius $r_m(z)$ = 2.7×10$^{-4}$+1.7×10$^{-4}$ ($z^2/f^2$-1), **g** the self-rotating non-diffracting ring bright lattices beam (NDABLB) with $\varphi_0(z)$ = -8.54$z$ rad and $\omega_0$ = -8.54 rad/m. The blue curves represent the propagating intensity on the vortex ring of these NDPVBs and the red arrow in **g** depicts the direction of self-rotation.

**New toolbox for Customizing NDPVB.**

Structured light beams[11], as solutions of the Helmholtz equation, can propagate in free space and have intriguing characteristic optical behaviors in propagation. For example, families of Gaussian modes[12] including Gaussian beam, Hermite–Gaussian beam, Laguerre–Gaussian beam, and Ince–Gaussian beam exhibit self-similar behavior, which means they hold scale-varying intensity patterns during propagation. Bessel beams, in contrast, are non-diffracting and have propagation-invariant and self-healing properties. Furthermore, the localized non-diffracting self-accelerating Airy beams[13] move along a curved parabolic trajectory in a way analogous to that of a projectile moving under the action of gravity[14].

Beyond these prototypical beams with confined optical behaviors, the area of maneuvering propagating behaviors of structured light beams has stimulated substantial research interest over the last decade, including arbitrary design of self-similar behaviors[15, 16], arbitrary transversal patterns of non-diffraction[17, 18], arbitrary trajectories of self-acceleration[19, 20] and so on. The current toolkit, however, is still in its infancy because each of the methods for carving out optical behaviors is based on a different theory and is incompatible with the others. Here we propose a generalized theory based on Fourier-space analysis to customize NDPVB at will so that it has controllable propagating intensity and phase while holding intriguing behaviors of self-accelerating, self-similar, and self-rotating. This goes beyond tailoring the transverse plane to the higher-dimensional propagating characteristics in structured light beams. The theoretical derivation is as follows.

A monochromatic light field propagating along the z-axis can be represented by its angular spectrum in cylindrical coordinates $(r, \varphi, z)$ as

$$U(r,\varphi,z) = (1/2\pi)^2 \iint A(k_r,\phi) e^{ik_r r \cos(\phi-\varphi)} e^{ik_z z} k_r dk_r d\phi \tag{5}$$

where $(k_r, \phi, k_z)$ represent the three-dimensional cylindrical coordinates in the Fourier-space (**k**-space), respectively, and $A(k_r, \phi)$ is the angular spectrum of the light field $U(r, \varphi, z = 0)$ in the focal plane at $z = 0$. The light field $U(r, \varphi, z)$ must obey the Helmholtz Equation by retaining the **k**-space relation $k_z = \sqrt{k^2 - k_r^2}$ while neglecting evanescent waves. The Fourier transformation expressed in equation (5) can be optically realized by a lens focusing process (shown in Fig. 1c-g): an incident light field distribution $A(k_r, \phi)$ in the front focal plane of the lens (focal length $f$) will be transformed into the focal field $U(r, \varphi, z=0)$ in the rear focal plane at $z = 0$, then yielding a light field distribution $U(r, \varphi, z)$ in an arbitrary plane at z after accounting for a free-space propagator $\exp(ik_z z)$ through equation (5). It should be noted that in the real-space coordinate system the radial wavenumber $k_r$ must be converted into the radial coordinate $\rho$ in the incident plane (i.e. the front focal plane) according to the relation $k_r = k\rho/f$.

A NDPVB can be factorized into the product of radial and azimuthal parts as $U(r, \varphi, z) = U(r, z)\exp(im\varphi)$, and so is for its angular spectrum at $z = 0$ in the form of $A(k_r, \phi) = A(k_r) \exp(im\phi)$. We get from equation (5) the following expression for the field distribution of the focal region by using the Bessel identity and $k_r = \sqrt{k^2 - k_z^2}$,

$$U(r,z) = 2\pi i^m \int_0^k A\left(\sqrt{k^2-k_z^2}\right) J_m(\sqrt{k^2-k_z^2} r) e^{ik_z z} k_z dk_z. \tag{6}$$

Equation (6) implies that a practical light field can be seen as a composition of ideal $m$th-order HOBBs with different complex weighting factors. We merely pay attention to the complex amplitude distribution on the NDPVB's vortex-ring, namely, $U(r = r_m, z) = C_m J_m(k_{rm} r_m)\exp(ik_{zm} z) = \exp(ik_{zm} z)$. The needed incident light field is calculated by the inverse transform of equation (6):

$$A\left(\sqrt{k^2-k_z^2},\phi\right) = \frac{e^{im\phi} \int_{-\infty}^{\infty} e^{ik_{zm}z} e^{-ik_z z} dz}{2\pi i^m rect\left(\frac{k_z}{2k}\right) J_m(\sqrt{k^2-k_z^2} r_m) k_z}, \tag{7}$$

where $rect(\ )$ represents the rectangle function. When $z = (-\infty, +\infty)$, $A\left(\sqrt{k^2-k_z^2},\phi\right) = \delta(k_z - k_{zm})\exp(im\phi)$ denotes the ideal NDPVB's angular spectrum (non-diffracting in an infinite range), which corresponds to an impractically thin incident field with a ring shape. When z is truncated within a constrained range like $rect(z/2f)$ in this work, $A\left(\sqrt{k^2 - k_z^2}, \phi\right)$ represents the angular spectrum of a quasi-NDPVB, which retains the properties of an ideal NDPVB in this range and is physically realizable. Based on the formalism of calculus, let us elucidate our recipe by revisiting the one-dimensional Fourier integral in equation (7); the integral can be regarded as the infinite sum of many field slices, each of which locates in the tiny z-axial range of (z, z+Δz) and is expressed by

$$A_z\left(\sqrt{k^2 - k_z^2}, \phi\right) = \frac{e^{ik_{zm}z} e^{-ik_z z} e^{im\phi}}{2\pi i^m rect\left(\frac{k_z}{2k}\right) J_m(\sqrt{k^2 - k_z^2} r_m) k_z}. \tag{8}$$

$A_z\left(\sqrt{k^2 - k_z^2}, \phi\right)$ represents the angular spectrum of a light sheet within (z, z+Δz). In this way, we can first sculpt all tiny light sheets and then accumulate them to reconstitute the integration to obtain the customized NDPVB.

**Customizing Propagating Intensity.** Although the NDPVBs are robust against disturbed turbid circumstances because of their non-diffracting and self-healing property, their intensity still tends to fade out due to the possible scattering or absorption in lossy media. Naturally, it becomes appealing to control and compensate for the intensity of these beams along the propagation direction. Let us assign an amplitude $\sqrt{I(z)}$ to the light sheet within (z, z+Δz) so that

$$A_z\left(\sqrt{k^2 - k_z^2}, \phi\right) = \frac{\sqrt{I(z)} e^{ik_{zm}z} e^{-ik_z z} e^{im\phi}}{2\pi i^m rect\left(\frac{k_z}{2k}\right) J_m(\sqrt{k^2 - k_z^2} r_m) k_z}, \tag{9}$$

Equation (9) can be envisioned as a means of controlling the intensity on the vortex ring of NDPVB in the single transverse plane at z. Consequently, the synthetic angular spectrum of the NDPVBs with shaped propagating intensity profiles can be calculated from the sum of all light sheets by

$$A\left(\sqrt{k^2 - k_z^2}, \phi\right) = \frac{e^{im\phi}}{2\pi i^m rect\left(\frac{k_z}{2k}\right) J_m(\sqrt{k^2 - k_z^2} r_m) k_z} \int_{-\infty}^{\infty} \sqrt{I(z)} e^{ik_{zm}z} e^{-ik_z z} dz \tag{10}$$

Our method in equation (10) can freely design the propagating intensity of NDPVB (for example, in the case of constant intensity $I(z) = rect(z/2f)$ with $f$ = 200 mm, $r_m$ = 150 um, and $m$ = 12, as illustrated in Fig. 1d), in contrast to the conventional methods like using the Gaussian ring aperture and a lens[8] to generate quasi-NDPVB (i.e. Bessel Gaussian beams) whose propagating intensity is not uniform, as shown in Fig. 1c. In Supplementary Experimental Demonstration 3, the corresponding experimental results for the uniform intensity $I(z) = rect(z/2f)$, the linearly increased intensity $I(z) = 0.5(-z/f+1)$, and exponentially increased intensity $I(z) = \exp(10z)/\exp(10f)$ along the z-direction are displayed. Another experiment in a particular application scenario where NDPVB propagates in lossy and scattering media like milk suspensions has also been conducted to demonstrate the viability of modifying the propagating intensity profiles. The intensity profile on the vortex ring along the z-direction can remain almost uniform by predesigning the propagating intensity of NDPVB to compensate for the attenuation of the media, even though the transversal total energy is also attenuating. The attenuation-compensated NDPVBs remaining the constant propagating intensity in different media will benefit a variety of optical applications under disturbed turbid circumstances, e.g. underwater and atmospheric communication, manipulation in some solutions, and imaging for biological samples.

**Customizing Propagating Phase.** Beam shaping plays a key role in optics and phase shaping is as crucial as intensity shaping. By imposing each light sheet within (z, z+Δz) with an extra phase $\delta(z)$, similar to the intensity adjusting in equation (9), the synthetic angular spectrum of the NDPVBs with shaped extra propagating phase on the vortex ring can be calculated from the sum of all light sheets by

$$A\left(\sqrt{k^2-k_z^2},\phi\right)=\frac{e^{im\phi}}{2\pi i^m rect\left(\frac{k_z}{2k}\right)J_m(\sqrt{k^2-k_z^2}r_m)k_z}\int_{-\infty}^{\infty}e^{i\delta(z)}e^{ik_{zm}z}e^{-ik_z z}dz. \qquad (11)$$

The total propagating phase of NDPVB after phase shaping is $k_{zm}z+\delta(z)$ on the vortex ring, and the inherent propagating phase $k_{zm}z$ is typically several orders of magnitude larger than $\delta(z)$, making it challenging to precisely observe the extra phase $\delta(z)$ in the scalar field. However, when $\delta(z)$ becomes the propagating phase difference of two polarization-based beams, the extra phase can be observed precisely by the propagating evolution of the polarization state's orientation in the synthetic vectorial field, because the polarization state's orientation is sensitive to phase difference in the vectorial field. This is beyond the scalar field scope of the current article and will be reported in another paper. Propagating phase shaping on the vortex ring not only paves the way to design vectorial NDPVBs but also plays an important role in the scalar field, e.g. when shaping the NDPVB with a Gouy phase of LGB, we can obtain a quasi-LGB possessing the properties of both LGB and customizable NDPVB.

**Customizing Self-accelerating Behaviors.** After Berry and Balazs theoretically predicted Airy beams in 1979, the self-accelerating beams have stimulated substantial research interest and have found a variety of applications, including particle manipulation and propelling[21], bending surface plasmons and electrons[22], curved plasma generation[23], light-sheet microscopy[24], single-molecule imaging[25], and others. There is little doubt that NDPVBs with arbitrarily self-accelerating capability provide more versatility and can also have significant advantages in many fields. According to the Fourier phase-shifting theorem, the light sheet within ($z$, $z+\Delta z$) will undergo a translation from ($x, y, z$) to ($x$-$g(z), y$-$h(z), z$) if a complex exponential $\exp(ik_x g(z)+ik_y h(z))$ is imposed on the angular spectrum so that it is explicitly expressed by

$$A_z\left(\sqrt{k^2-k_z^2},\phi\right)=\frac{e^{ik_x g(z)+ik_y h(z)+ik_{zm}z}e^{-ik_z z}e^{im\phi}}{2\pi i^m rect\left(\frac{k_z}{2k}\right)J_m(\sqrt{k^2-k_z^2}r_m)k_z}, \qquad (12)$$

where ($k_x, k_y$) is the transverse wavenumber in Cartesian coordinates with $k_x = k_r\cos\phi$ and $k_y = k_r\sin\phi$. Equation (12) can be envisioned as a means for controlling the position of NDPVB in the single transverse plane at $z$. Consequently, the synthetic angular spectrum of the self-accelerating NDPVB can be calculated from the sum of all light sheets by

$$A\left(\sqrt{k^2-k_z^2},\phi\right)=\frac{e^{im\phi}}{2\pi i^m rect\left(\frac{k_z}{2k}\right)J_m(\sqrt{k^2-k_z^2}r_m)k_z}\int_{-\infty}^{\infty}e^{ik_x g(z)+ik_y h(z)+ik_{zm}z}e^{-ik_z z}dz. \qquad (13)$$

And the self-accelerating trajectories can also be expressed in terms of the displacement vector $\mathbf{s}(z)=g(z)\hat{x}+h(z)\hat{y}=(g(z),h(z))$ with $\hat{x}$ and $\hat{y}$ denoting the unit vectors of the $x$ and $y$ directions, respectively. The self-accelerating parabolic NDPVB behaving like an Airy beam is shown in Fig. 1e with the same parameters as Fig. 1d but $\mathbf{s}(z) = 1.8\times10^{-4}$ [0, 1-$z^2/f^2$]. Self-accelerating NDPVB can act as a snowblower conveying particles from one compartment to another[26] or convey information to the obscured invisible receiver[27], which has been experimentally verified in Supplementary Experimental Demonstration 4.

**Customizing Self-similar Evolution.** A fascinating query naturally arises after revisiting equations (2-3): Can the NDPVB be tailored to change in scale like the self-similar LGB? This idea can be realized using the versatile angular spectrum theory. By accounting for the z-dependent radius in equation (3), the corresponding radial and longitudinal components of the wavevector of NDPVB can be rewritten as

$$k_{rm}(z)=\left(C_1\times 2m\sqrt{1+\frac{1}{m}}+C_2\right)\!\bigg/r_m(z) \qquad (14\text{-}1)$$

$$k_{zm}(z)=\sqrt{k_0^2-k_{rm}^2(z)}. \qquad (14\text{-}2)$$

Assigning the light sheet of NDPVB within $(z, z+\Delta z)$ with a radius $r_m(z)$, we get

$$A_z\left(\sqrt{k^2-k_z^2},\phi\right)=\frac{\exp(i\int_{-\infty}^{z}k_{zm}(z)dz)e^{-ik_z z}e^{im\phi}}{2\pi i^m rect\left(\frac{k_z}{2k}\right)J_m(\sqrt{k^2-k_z^2}r_m)k_z}, \quad (15)$$

Consequently, the synthetic angular spectrum of the self-similar NDPVBs can be calculated from the sum of all light sheets by

$$A\left(\sqrt{k^2-k_z^2},\phi\right)=\frac{e^{im\phi}}{2\pi i^m rect\left(\frac{k_z}{2k}\right)J_m(\sqrt{k^2-k_z^2}r_m)k_z}\int_{-\infty}^{\infty}\exp(i\int_{-\infty}^{z}k_{zm}(z)dz)e^{-ik_z z}dz. \quad (16)$$

The self-similar NDPVB behaving like a LGB beam is shown in Fig. 1f, having the same parameters as Fig. 1d but $r_m(z) = 2.7\times10^{-4}+1.7\times10^{-4}\,(z^2/f^2-1)$. Supplementary Experimental Demonstration 4 demonstrates how the self-similar NDPVB can transmit data to the hidden, obscured receiver as well. These self-accelerating and self-similar NDPVBs with controllable propagating intensity and phase enrich the "toolkit" of beam shaping and can be utilized in a variety of applications such as filamentation, particle manipulation, biomedical imaging, plasmons, and material processing, among others.

Tailoring the self-similar behavior by equation (16) and shaping the extra propagating phase $\delta(z)$ by equation (11) are both achieved by controlling the propagating phase along the z-direction, thus yielding a modified longitudinal wavevector expressed as

$$k_{zm1}(z)=\frac{d(\int_{-\infty}^{z}k_{zm}(z)dz+\delta(z))}{dz}=k_{zm}(z)+\delta'(z), \quad (17)$$

and the actual radius is given by

$$r_m(z)=\left[C_1\left(2|m|\sqrt{1+1/|m|}\right)+C_2\right]\Big/\sqrt{k_0^2-(k_{zm}(z)+\delta'(z))^2}. \quad (18)$$

Because, for synthesized vectorial fields, $\delta'(z)$ is typically several orders of magnitude smaller than the longitudinal wavevector component $k_{zm}(z)$, equation (18) suggests that the variation of vortex radius resulting from the phase modification $\delta(z)$ can be neglected.

**Customizing Self-rotating Behaviors.** The superposition of two NDPVBs that differ in both their linear ($k_{zm}$) and angular ($m$) momenta from equations (3-4) would result in a standing-wave pattern in both the longitudinal and azimuthal directions, thereby realizing a spiraling intensity pattern. For example, the three-dimensional intensity pattern on the vortex ring of two coaxially propagating NDPVBs with TCs $m$ and $n$ respectively can be expressed as

$$\begin{aligned}I(r,\varphi,z)&=\left|C_m J_m(k_{rm}r)e^{im\varphi}e^{ik_{zm}z}+C_n J_n(k_{rn}r)e^{in\varphi}e^{ik_{zn}z}\right|^2\\&=(C_m J_m(k_{rm}r))^2+(C_n J_n(k_{rn}r))^2+2C_m C_n J_m(k_{rm}r)J_n(k_{rn}r)\cos((m-n)(\varphi+\frac{k_{zm}-k_{zn}}{m-n}z))\\&=I(r,\varphi+\varphi_0(z))\end{aligned} \quad (19)$$

Equation (19) shows that this structured NDPVB is also non-diffracting but exhibits the self-rotating behavior upon propagation with the angular velocity $\omega_0 = d\varphi_0(z)/dz = (k_{zm}-k_{zn})/(m-n)$, signed depending on the clockwise or counterclockwise direction of rotation. For the vortex ring at $r = r_0$, the self-rotating intensity pattern is given by $I(r = r_0, \varphi, z) = 2+2\cos((m-n)(\varphi+\varphi_0(z)))$, in which there are $N = |m-n|$ bright spots produced by constructive interference on the vortex ring with an equal number of dark spots by destructive interference, forming an annular optical lattice[28]. Such an optical lattice has an intensity gradient around the bright places, which can be modulated by adjusting $N$ and $r_0$. We refer to this self-rotating structured NDPVB as a non-diffracting annular bright lattice

beam(NDABLB), which can be beneficial for traping and guiding batches of particles with large refractive index and atoms [29,30,31]. A self-rotating NDABLB with parameters ($r_0$ = 411.4 um, $m$ = 12, $n$ = 17) has $N$ = 5 and $\omega_0$ = -8.54 rad/m, as shown in Fig. 1g, which is consistent with simulation and is further verified in Supplementary Experimental Demonstration 5.

Additionally, optically trapping and guiding low-index particles, whose refractive index is lower than the surrounding medium, is of great interest in biology and medicine. For example, a drug can be enclosed in trapped microbubbles and then moved to target tissues to observe the reaction of target cells to the drug[32]. By slightly tuning the radii of two NDPVBs in the superposition, a self-rotating non-diffracting annular dark lattice beam(NDADLB) with the same numbers $N=|m-n|$ and angular velocity $\omega_0 = (k_{zm}-k_{zn})/(m-n)$ is produced and also demonstrated in Supplementary Experimental Demonstration 5 where each dark trap represents an optical vortex with TC = 1 or -1 (the basic formalism is described in ref.[33]). This beam is competent for trapping and guiding low-index particles.

The multidimensionally customizable structured NDPVBs such as NDABLBs and NDADLBs can serve as a more potent optical Archimedes' screw[34, 35] with customizable trajectories for transporting trapped particles in batches over extended distances either downstream (guiding particles) or upstream (tracking particles) from the direction of momentum flow. Furthermore, the self-rotating behavior with an adjustable z-dependent angular velocity can also be realized through the phase shaping according to $\varphi_0(z) = [(k_{zm}(z)-k_{zn}(z))z+(\delta_m(z)-\delta_n(z))]/(m-n)$ and $\omega(z) = d\varphi_0(z)/dz$, where $k_{zm}(z)$ and $k_{zn}(z)$ are defined from equation (14) of $m$th- and $n$th- order NDPVBs with $\delta_m(z)$ and $\delta_n(z)$ representing the additional propagating phases.

**Customizing Piecewise Beam.** Based on the formalism of calculus, our scheme can also configure piecewise NDPVBs for the parameters $I(z)$, $\delta(z)$, $\mathbf{s}(z)$, $r_m(z)$, and even $m(z)$. Imparting different TC $m(z)$ to each light sheet so that the synthetic angular spectrum is given by

$$A_m\left(\sqrt{k^2-k_z^2},\phi\right)=\frac{1}{2\pi rect\left(\frac{k_z}{2k}\right)k_z}\int_{-\infty}^{\infty}\frac{e^{ik_{zm(z)}z}e^{im(z)\phi}e^{-ik_z z}}{i^{m(z)}J_{m(z)}(\sqrt{k^2-k_z^2}r_{m(z)})}dz \qquad (20)$$

For example, for the TC-customized NDPVB with $m(z) = -12rect(z/f+0.5)+17rect(z/f-0.5)$ meaning that $m$ = -12 in [-200 mm, 0 mm] and $m$ = 17 in [0 mm, 200 mm], its complex field distribution, shown on the left of Moive S6, indicates the TC of vortex ring switches from -12 to 17 at $z$ = 0 mm plane. We have further confirmed by the right panel of Movie S6 that the TC changes in a self-rotating NDABLB produced by the coaxial superposition of the above NDPVB and the NDPVB with TC = 8; the number of the resulting bright lattices decreases from 20 to 9 at the $z$ = 0 plane, accompanied by a self-rotation direction reverse. This demonstrates that we can manage the local OAM's propagating evolution on the vortex ring (the region of interest of NDPVB). Other behaviors can also be piecewise shaped, such as how a customized NDPVB can imitate Gaussian modes (self-similar), Bessel modes (non-diffracting, self-healing), and Airy modes (self-accelerating) at different propagating distances.

As demonstrated above, our Fourier-space design enables piecewise beam shaping and controls propagating properties of beams, such as the propagating intensity $I(z)$ and phase $\delta(z)$, the self-accelerating behavior represented by $\mathbf{s}(z)=(g(z), h(z))$, and the self-similar behavior represented by $r_m(z)$ and the self-rotating behavior represented by $\varphi_0(z)$. The proposed NDPVB's multidimensional and customizable properties make it an appealing and promising candidate for conventional vortex beams with potential applications ranging from imaging, microscopy, optical communication, metrology, and quantum information processing to light-matter interactions. By the way, this versatile shaping toolbox also works for 0th-order Bessel beams when $m$ = 0.

## The internal flows of customized NDPVBs.

The behavior shaping of NDPVBs deals with their exterior performances: a beam field is described as it appears to an 'external' observer. The internal energy flows (or momentum density), which underlies beam structure formation and behavior transformation during propagation, on the other hand, provide an efficient and natural way for peering into light beam fields and studying their most intimate and deep features. We performed the following analysis on the momentum density of customized NDPVBs, for facilitating our understanding of how our design scheme of propagating properties manipulates the internal flows.

Provided monochromatic scalar paraxial beam in free space, the momentum density can be expressed as the intensity-weighted wavevector density[36], $\boldsymbol{p}(x, y, z) = I(x, y, z) \boldsymbol{k}(x, y, z)$, where the wavevector density $\boldsymbol{k}(x, y, z)$ is defined as the gradient of the phase distribution $\Phi(x, y, z)$ expressed as

$$\begin{aligned}\boldsymbol{k}(x,y,z) &= \frac{\partial \Phi(x,y,z)}{\partial x}\hat{\boldsymbol{x}} + \frac{\partial \Phi(x,y,z)}{\partial x}\hat{\boldsymbol{y}} + \frac{\partial \Phi(x,y,z)}{\partial x}\hat{\boldsymbol{z}} \\ &= k_x(x,y,z)\hat{\boldsymbol{x}} + k_y(x,y,z)\hat{\boldsymbol{y}} + k_z(x,y,z)\hat{\boldsymbol{z}},\end{aligned} \tag{21-1}$$

and

$$\begin{aligned}\boldsymbol{k}(x,y,z) &= \frac{\partial \Phi(x,y,z)}{\partial r}\hat{\boldsymbol{r}} + \frac{1}{r}\frac{\partial \Phi(x,y,z)}{\partial \varphi}\hat{\boldsymbol{\varphi}} + \frac{\partial \Phi(x,y,z)}{\partial z}\hat{\boldsymbol{z}} \\ &= k_r(x,y,z)\hat{\boldsymbol{r}} + k_\varphi(x,y,z)\hat{\boldsymbol{\varphi}} + k_z(x,y,z)\hat{\boldsymbol{z}},\end{aligned} \tag{21-2}$$

where $\hat{\boldsymbol{r}}, \hat{\boldsymbol{\varphi}}, \hat{\boldsymbol{z}}$ are the unit vectors in cylindrical coordinates. According to the continuity equation[37]. The transverse wavevector/momentum density governs how the intensity structure of beams evolute and is given by

$$\boldsymbol{k}_\perp(x,y,z) = k_x(x,y,z)\hat{\boldsymbol{x}} + k_y(x,y,z)\hat{\boldsymbol{y}} = k_r(x,y,z)\hat{\boldsymbol{r}} + k_\varphi(x,y,z)\hat{\boldsymbol{\varphi}}. \tag{22}$$

The spiral phase in equation (4) shows that the ideal NDPVBs $U(x, y, z)$ have an inherent azimuthal wavevector,

$$k_\varphi(x,y,z) = \frac{1}{r}\frac{\partial(m\varphi)}{\partial \varphi} = \frac{m}{r} = \frac{m}{y\sin\theta + x\cos\theta}, \tag{23}$$

where $\theta$ is the polar angle. Changing the TC $m$ of an NDPVB having a constant radius $r_0$ is equivalent to adjusting the azimuthal wavevector $m/r_0$ on the vortex ring. When shaping the self-accelerating behavior of a NDPVB in the form of $U(x-g(z), y-h(z), z)$, it will induce an additional transversal wavevector on the vortex ring written in Cartesian coordinates as

$$\Delta \boldsymbol{k}_\perp(x,y,z) = k_0 g'(z)\hat{\boldsymbol{x}} + k_0 h'(z)\hat{\boldsymbol{y}} = k_0 \boldsymbol{s}'(z), \tag{24}$$

where a single prime symbol denoting the first-order derivative with respect to $z$ (the full derivation can be found in Supplementary theory 4). The self-accelerating behavior will also have an effect on the inherent azimuthal wavevector density in equation (23) but only changes the center from (0, 0) to ($g(z), h(z)$), given by

$$k_\varphi(x,y,z) = \frac{m}{(y-h(z))\sin\theta + (x-g(z))\cos\theta}. \tag{25}$$

Shaping a self-similar NDPVB having a $z$-dependent radius $r_m(z)$ and its complex distribution, $U(r = r_m(z_0), \varphi, z_0)$, at the initial $z = z_0$ plane, will induce another additional radial wavevector density, expressed as (the full derivation can be found in Supplementary theory 4 as well)

$$\Delta k_r(x,y,z) = k_0 \frac{r'_m(z)}{r_m(z_0)} r. \tag{26}$$

The self-similar behavior will also affect the inherent azimuthal wavevector density as

$$k_\varphi(x,y,z) = \frac{m}{r(z)} = \frac{r_m(z_0)}{r_m(z)}\frac{m}{y\sin\theta + x\cos\theta}. \tag{27}$$

In summary, the total transversal wavevector density of customized NDPVBs with behaviors of self-accelerating and self-similar can be expressed as

$$\begin{aligned}\boldsymbol{k}_\perp(x,y,z) &= k_\varphi(x,y,z)\hat{\boldsymbol{\varphi}} + \Delta\boldsymbol{k}_\perp(z) + \Delta k_r(x,y,z)\hat{\boldsymbol{r}} \\ &= \frac{r_m(z_0)m}{r_m(z)(y-h(z))\sin\theta + (x-g(z))\cos\theta}\hat{\boldsymbol{\varphi}} + k_0 g'(z)\hat{\boldsymbol{x}} + k_0 h'(z)\hat{\boldsymbol{y}} \\ &\quad + k_0 \frac{r'_m(z)}{r_m(z_0)}(y-h(z))\sin\theta + (x-g(z))\cos\theta\,\hat{\boldsymbol{r}},\end{aligned} \quad (28)$$

Considering the place on vortex ring, the most noteworthy and the region of interest of NDPVB, that satisfies

$$((y-h(z))\sin\theta)^2 + ((x-g(z))\cos\theta)^2 = r_m^2(z), \quad (29)$$

the total transversal wavevector density can be written as

$$\boldsymbol{k}_\perp(r_m(z),\varphi,z) = \frac{m}{r_m(z)}\hat{\boldsymbol{\varphi}} + k_0 g'(z)\hat{\boldsymbol{x}} + k_0 h'(z)\hat{\boldsymbol{y}} + k_0 r'_m(z)\hat{\boldsymbol{r}}, \quad (30)$$

and thus the transversal momentum density on the vortex ring is given by

$$\boldsymbol{p}_\perp(r_m(z),\varphi,z) = I(z)\left(\frac{m}{r_m(z)}\hat{\boldsymbol{\varphi}} + k_0 g'(z)\hat{\boldsymbol{x}} + k_0 h'(z)\hat{\boldsymbol{y}} + k_0 r'_m(z)\hat{\boldsymbol{r}}\right). \quad (31)$$

The transversal momentum density of customized NDPVBs is also calculated by the complex field distributions, following from the cycle-average Poynting vector[37] given as $\boldsymbol{p} = \varepsilon_0/(2\omega)\operatorname{Im}\left[\boldsymbol{U}^* \times (\nabla \times \boldsymbol{U})\right]$, where $\varepsilon_0$ and $\omega$ are the vacuum permittivity and circular frequency, respectively, and the calculated results can be found in Supplementary Demonstration of Internal Flows, which are well consistent with the prediction of equations (31).

Equation (31) shows the relationship between the customized parameters ($m$, $I(z)$, $g(z)$, $h(z)$, $r_m(z)$) and the real-space internal flows, suggesting that multi-modal customization is associated with reforming internal flows, e.g. the self-accelerating behavior $s(z) = (g(z), h(z))$ is realized by inducing the internal flow ($I(z)k_0 g'(z)$, $I(z)k_0 h'(z)$) in the $x$ and $y$ dimensions and the self-similar behavior $r_m(z)$ is realized by inducing the internal flow $I(z)k_0 r'_m(z)$ in the radial dimension, while both behaviors affect the inherent azimuthal flow by coordinates translating and rescaling.

## Conclusion

Many applications, including fiber optic data transmission, spatial OAM mode (de)multiplexing communication, and particle manipulation, are hampered by the obstinate and rigid characteristics of traditional VBs, such as LGB and HOBB, whose radii inevitably grow up with OAM. This inherent limitation of traditional VBs is completely overcome by the proposed NDPVB, whose vortex radius can be arbitrarily accurately designed regardless of the OAM carried when propagating perfectly without divergence and possessing properties of non-diffracting, self-healing. To make NDPVBs more versatile and robust, we work out a toolkit to multidimensionally customize NDPVB at will so that it has controllable propagating intensity and phase, resulting in intriguing customizable behaviors like self-accelerating, self-similar, and self-rotating. The light field flow analysis revealed and confirmed that the multidimensional customization of NDPVBs is accomplished by inducing corresponding multidimensional internal flows, unveiling the nature of structure formation and behavior transformation of structured light beams. The results of optical experiments, including the attenuation-compensated experiments in lossy media and the self-healing, self-accelerating, self-similar, and self-rotating experiments, sufficiently show the versatility and malleability of these multidimensionally customizable NDPVBs. We believe this toolkit will make NDPVBs an ideal candidate of VBs for potential applications in fields like imaging, microscopy, optical communication, metrology, quantum information processing, and light-matter interactions.

**Acknowledgements**

This work was supported by National Key R&D Program of China (2018YFA0306200 and 2017YFA0303700); National Natural Science Foundation of China (91750202 and 11922406).

# Supplementary Information for

# Versatile Non-diffracting Perfect Vortex Beam

Wenxiang Yan,[1] Yuan Gao,[1] Zheng Yuan,[1] Zhe Weng,[1] Zhi-Cheng Ren,[1] Xi-Lin Wang,[1] Jianping Ding,[1,2,3] and Hui-Tian Wang[1]

[1]Nanjing University, National Laboratory of Solid Microstructures and School of Physics, Nanjing, China

[2]Nanjing University, Collaborative Innovation Center of Advanced Microstructures, Nanjing, China

[3]Nanjing University, Collaborative Innovation Center of Solid-State Lighting and Energy-Saving Electronics, Nanjing, China

Correspondence to: jpding@nju.edu.cn

**Supplementary theory:**
1. The derivation of equation(2): the solution of vortex radius of HOBB;
2. The derivation of the normalized amplitude coefficient in equation (4);
3. Basics of LGB shown in Fig. 1a;
4. Analyzing the internal flows of customized NDPVBs.

**Supplementary Experimental Method.**

**Supplementary Experimental Demonstration:**
1. The perfection of non-diffracting NDPVB;
2. The self-healing behavior of NDPVB;
3. NDPVBs with customized propagating intensity in air and lossy media;
4. The self-accelerating and self-similar behavior of customized NDPVB;
5. The self-rotating behavior and optical ring lattice of structured NDPVBs.

**Supplementary Demonstration of Internal Flows.**

**Supplementary Movies: Movies S1-S15 .**

**Supplementary Reference.**

## Supplementary theory

### 1. The derivation of equation(2): the solution of vortex radius of HOBB.

The field distribution of an $m$th-order Bessel beam in cylindrical coordinates $\mathbf{r}=(r,\varphi,z)$ is given by

$$U(r,\varphi,z)=J_m(k_{r0}r)e^{im\varphi}e^{ik_{z0}z}. \tag{S1}$$

The radius of HOBBs' vortex where the maximum intensity exhibits is determined through the smallest non-zero solution of $d(J_m(k_{r0}r))/dr=0$ with $r>0$ and $m\geq1$. Hence, we need to find the minimum non-zero extreme point of the $m$th-order Bessel function $J_m(x)$. The distributions of Bessel functions $J_m(x)$ with $m$ = 12, 17, 22, and 26 are displayed in Fig. S1a and the maximum positions increase with the order $m$, indicating that the vortex radius of HOBB grows up with the topological charge.

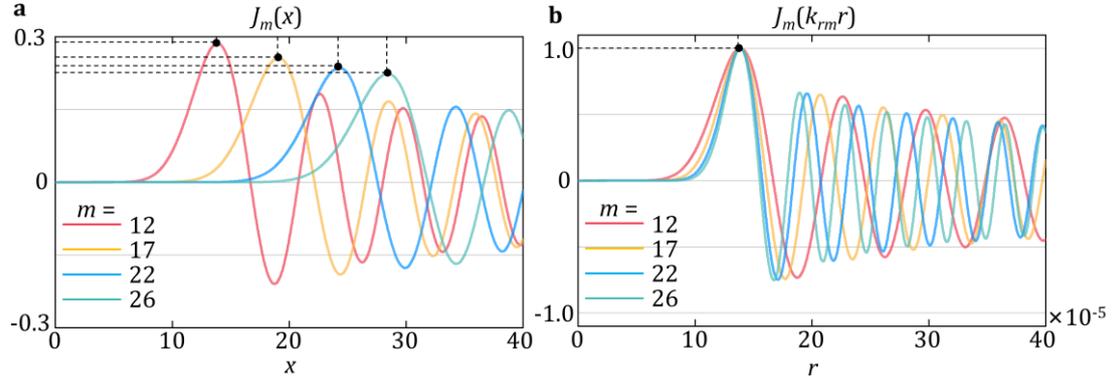

Fig. S1 **a** the distributions of Bessel functions $J_m(x)$ with $m$ = 12, 17, 22, 26. **b** the amplitude distributions of the NDPVBs generated by equation (2) and the normalized amplitude coefficient in equation (S10) with $m$ = 12, 17, 22, 26 but the same radius $x_0$=13.88. The black points in both **a** and **b** are the positions where minimum non-zero extreme points exist.

According to equation (5.2.6) of ref.[1] for the Bessel function, $d(J_m(x))/dx=0$ can be rewritten as $d(J_m(x))/dx=(J_{m-1}(x)-J_{m+1}(x))/2=0$. Although Bessel functions are infinitely extending functions with infinite extreme points, we restrict attention to the minimum non-zero extreme point whose position is close to the origin. Using the asymptotic formula (5.2.6) of ref.[1], $J_m(x)\approx x^m/(2^m m!)$, for small values of $x$ that meets our requirement, we get

$$\frac{x^{m-1}}{2^{m-1}(m-1)!}-\frac{x^{m+1}}{2^{m+1}(m+1)!}=0. \tag{S2}$$

The approximate solution of the minimum non-zero extreme point of the $m$th order Bessel function becomes

$$x_m \approx 2m\sqrt{1+\frac{1}{m}}. \tag{S3}$$

A comparison reveals a deviation between with the minimum non-zero extreme point of $J_m(x)$ and the approximated value in equation (S3). After a non-linear regression fitting, the more precise location of the minimum non-zero extreme point is found to be

$$x_m \approx 2C_1 m\sqrt{1+\frac{1}{m}}+C_2. \tag{S4}$$

where $C_1$ and $C_2$ are the two correction parameters determined as follows

$$\begin{cases} C_1=0.5207 \text{ and } C_2=0.7730 \text{ for } m\leq 40 \\ C_1=0.5058 \text{ and } C_2=2.0223 \text{ for } m>40 \end{cases} \tag{S5}$$

In Fig. S2, the result from equation (S4) (the red-dot curve) is in excellent agreement with the actual minimum non-zero extreme location (the black curve); the percentage error (the blue curve) indicates that: when $m>4$, the error is less than 1%; when $m\geq 20$, the error is always less than 0.5% and finally remaining less than 0.1% with the larger $m$. When compared to the vortex ring's relatively large width, this error is typically insignificant and thus can be ignored.

Since the sign of the order of Bessel functions does not affect their intensity distribution, equation (2) in the main text concerning the vortex radius of HOBBs follows immediately, and so we get equation (3) for calculating $k_{rm}$ of chosen radius $r_m$ and order $m$. Equation (3) can be used to arbitrarily tailor the vortex radius of HOBB on-demand, resulting in TC-independent HOBBs, i.e. NDPVBs, some amplitude distributions of which are shown in Fig. S1b.

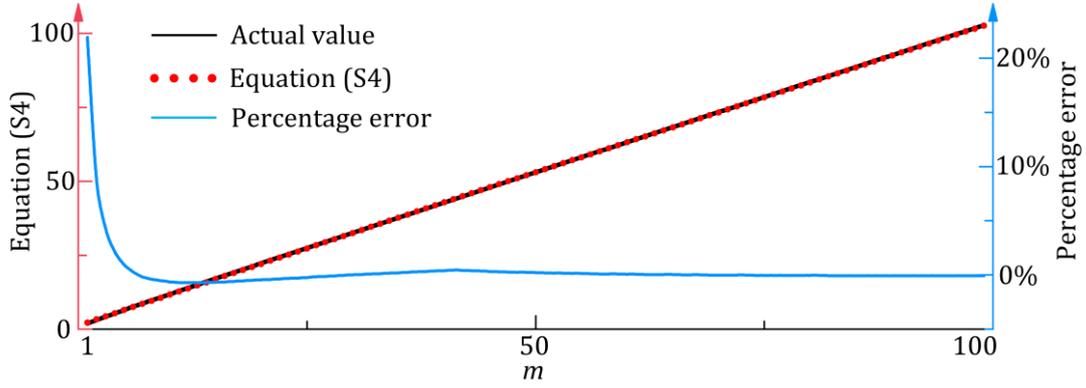

Fig. S2. Comparison of the results from equation (S4) (the red-dot curve) and the actual location of the minimum non-zero extreme point of the $m$th-order Bessel function (the black curve). The blue curve represents the percentage error that is defined by the ratio of the error over the actual value.

## 2. The derivation of the normalized amplitude coefficient in equation (4).

The maximum value of the $m$th-order Bessel function can be approximated by [2]

$$max_1(J_m) \approx \sqrt[3]{2/m}\, Ai(-0.809/\sqrt[3]{2}) = 0.632/\sqrt[3]{m}, \quad (S6)$$

where $Ai$ is the Airy function. Fig. S3a illustrates the accuracy of equation (S6); the results from equation (S6) (the red-dot curve in Fig. S3a) show a deviation trend when compared with the real value of the Bessel functions' maxima (the black curve), and the percentage error (the blue curve) becomes larger than 6% as $m$ increases. A non-linear regression fitting results in the following corrected relation of Bessel functions' maxima versus $m$,

$$max_2(J_m) = 1/\sqrt[3]{C_3 m + C_4} \quad (S7)$$

where $C_3 = 3.2823$, $C_4 = 4.0849$. In Fig. S3b, the results from equation (S7) (the red-dot curve) are in great agreement with the real look-up values (the black curve) and the percentage error (the blue curve) indicates that: when $m>2$, the error is less than 5%; when $m>8$, the error is always less than 1% and finally remaining less than 0.1% when $m>21$. Hence, the normalized amplitude coefficient is expressed as

$$C_m^{Amp} = 1/max_2(J_m) = \sqrt[3]{3.2823|m| + 4.0849}. \quad (S8)$$

When combining equation (3) and equation (S8), we can generate the NDPVB in equation (4) whose vortex radius and the maximum amplitude are the same regardless of the topological charge $m$ shown in Fig. S1b.

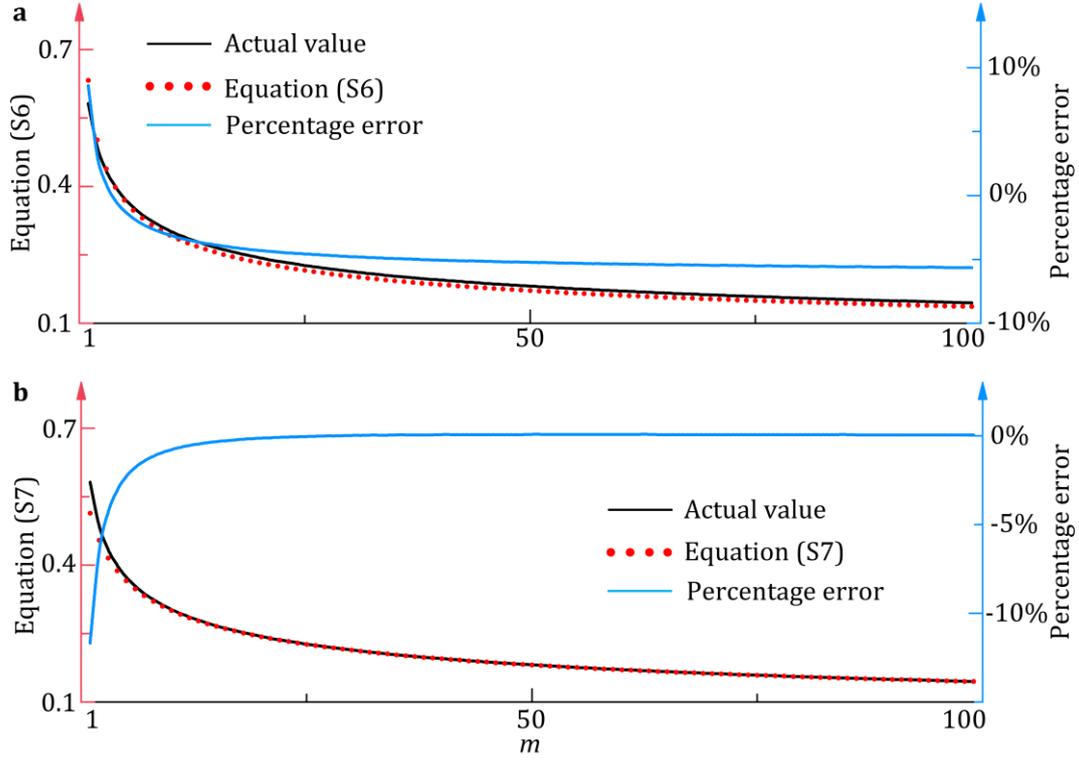

Fig. S3. **a** Comparison of the maximum values calculated from equation (S6) and the actual Bessel functions. The black, red-dot, and blue curves represent maximum values from equation (S6) and the Bessel functions, as well as the percentage error between them, respectively. **b** Same as **a** for the maximum values but calculated from equation (S7).

### 3. Basics of LGB shown in Fig. 1a.

The complex amplitude distribution of a LGB in cylindrical coordinates $(r, \varphi, z)$ is represented by

$$U_{m,p}(r,\varphi,z) = \left[\frac{W_0}{W(z)}\right]\left[\frac{r}{W(z)}\right]^m L_p^m\left[\frac{2r^2}{W^2(z)}\right]\exp\left(-\frac{r^2}{W^2(z)}\right) \\ \times \exp\left[ik_0 z + ik\frac{r^2}{2R(z)} + im\varphi - i(m+2p+1)\zeta(z)\right], \tag{S9-1}$$

where $L_p^m$ denotes the Laguerre polynomial with the azimuthal index (TC) $m$ and the radial index $p$, and $k_0 = 2\pi/\lambda$ is the wavenumber with $\lambda$ being the wavelength, and

$$W(z) = W_0\sqrt{1+\left(\frac{z}{z_0}\right)^2},\ R(z) = z\left[1+\left(\frac{z_0}{z}\right)^2\right],\ \zeta(z) = \tan^{-1}\left(\frac{z}{z_0}\right),\ z_0 = \frac{\pi W_0^2}{\lambda},\ W_0 = \sqrt{\frac{\lambda z_0}{\pi}}. \tag{S9-2}$$

with $W_0$ denoting the beam waist.

### 4. Analyzing the internal flows of customized NDPVBs.

If a NDPVB is propagating paraxially in the $y$-$z$ plane with trajectory $\boldsymbol{s}(z)=(0, h(z))$ at a constant angle $\alpha_y$, there must exist an extra exponential $\exp(ik_0\sin\alpha_y y)$ with a constant wavevector component $k_y = k_0\sin\alpha_y \approx k_0 h'(z)$ to dominate this evolution (such as being deflected by a prism). If the propagating angle varies with $z$ as expressed by $\alpha_y(z)$, the equivalent extra exponential is $\exp(ik_0\sin\alpha_y(z)y)$ with self-evolving wavevector component

$k_y(z)=k_0\sin\alpha_y(z)\approx k_0h'(z)$ along the z-direction. Similarly, when the NDPVB is self-accelerating in the x-z plane $s(z)=(g(z), 0)$, the extra wavevector component is $k_x(z)=k_0g'(z)$. Therefore, we speculate that for self-accelerating NDPVB $U(x-g(z), y-h(z), z)$ with $s(z)=(g(z), h(z))$, there will exhibit an additional self-evolving transversal wavevector $k_0s'(z)=(k_0g'(z), k_0h'(z))$.

The complex distribution of NDPVB at the initial $z=z_0$ plane is expressed in cylindrical coordinates as $U(r, \varphi, z_0)$, which can be transformed to the expression in Cartesian coordinates as

$$U(r,\varphi,z_0)=U(x,y,z_0)=U(-\varphi\sin\theta+r\cos\theta,\varphi\cos\theta+r\sin\theta,z_0). \tag{S10}$$

The complex distribution of the self-similar NDPVB $U(r(z), \varphi, z)$ with a z-dependent radius $r_m(z)$ at arbitrary z plane ($z>z_0$) can be also converted to the expression in Cartesian coordinates as

$$U(r(z),\varphi,z)=U(x(z),y(z),z)=U(-\varphi\sin\theta+r(z)\cos\theta,\varphi\cos\theta+r(z)\sin\theta,z). \tag{S11}$$

The self-similar NDPVB propagates with the changing radial scale while maintaining intensity pattern so that $r(z)=(r_m(z)/r_m(z_0))r$. Unlike the self-accelerating NDPVB $U(x-g(z), y-h(z), z)$ which traces out the same trajectory $s(z)=(g(z), h(z))$ at each position $(x, y)$ due to its global phase with a constant gradient $k_0s'(z)$, the self-similar NDPVBs $U(x(z), y(z), z)$ can be perceived as undergoing its own self-accelerating trajectory at each position $(x, y)$ according to

$$\boldsymbol{s}(x,y,z)=\Delta x(x,y,z)\hat{\boldsymbol{x}}+\Delta y(x,y,z)\hat{\boldsymbol{y}}, \tag{S12-1}$$

where

$$\Delta x(x,y,z)=x(z)-x=\left(\frac{r_m(z)}{r_m(z_0)}-1\right)r\cos\theta \tag{S12-2}$$

$$\Delta y(x,y,z)=y(z)-y=\left(\frac{r_m(z)}{r_m(z_0)}-1\right)r\sin\theta \tag{S12-3}$$

The wavevector analysis of the self-accelerating NDPVB $U(x-g(z), y-h(z), z)$ shows that there must be an additional local transversal wavevector density $\Delta\boldsymbol{k}_\perp(x,y,z)$ expressed by

$$\Delta\boldsymbol{k}_\perp(x,y,z)=k_x(x,y,z)\hat{\boldsymbol{x}}+k_y(x,y,z)\hat{\boldsymbol{y}}, \tag{S13-1}$$

where

$$k_x(x,y,z)=k_0\frac{\partial\Delta x(x,y,z)}{\partial z}=k_0\frac{r'_m(z)}{r_m(z_0)}r\cos\theta \tag{S13-2}$$

$$k_y(x,y,z)=k_0\frac{\partial\Delta y(x,y,z)}{\partial z}=k_0\frac{r'_m(z)}{r_m(z_0)}r\sin\theta, \tag{S13-3}$$

Equation (S13) can be rewritten as

$$\Delta\boldsymbol{k}_\perp(x,y,z)=k_r(x,y,z)\hat{\boldsymbol{r}}=k_0\frac{r'_m(z)}{r_m(z_0)}r\hat{\boldsymbol{r}}. \tag{S14}$$

Therefore, when shaping the self-similar NDPVB $U(r(z), \varphi, z)$ with a z-dependent radius $r_m(z)$, there will exhibit another additional radial wavevector component $k_r(x, y, z)$. To be more specific, we can derive the expression of the corresponding phase $\Phi(x, y, z)$ with the local phase gradient $(k_x(x,y,z), k_y(x,y,z))$ from the relationship, $\partial\Phi(x,y,z)/\partial x=k_x(x,y,z)$ and $\partial\Phi(x,y,z)/\partial y=k_y(x,y,z)$, given by

$$\Phi(x,y,z)=\frac{k_0}{2}\frac{r'_m(z)}{r_m(z_0)}(y\sin\theta+x\cos\theta)^2=\frac{k_0}{2}\frac{r'_m(z)}{r_m(z_0)}r^2, \tag{S15}$$

which represents a self-evolving radial quadratic phase. Comparing with the quadratic phase of lens with the focal length $f$, $-\pi r^2/(\lambda f)$, equation (S15) can be thought of as a "lens" with self-evolving focal length $f(z)= -r_m(z_0)/r'_m(z)$. For instance, for a self-similar and linearly stretching NDPVB, i.e. $r_m(z)= r_m(z_0)+Dz$ with $r'_m(z)=D>0$, equation (S15) resembles the phase of a concave lens. When the self-similar NDPVB is linearly shrinking with $r'_m(z)=D<0$, equation

(S15) depicts the phase of a convex lens. It is simple to demonstrate that the corresponding concave and convex lenses can linearly stretch and contract the NDPVB/HOBB in accordance with the predetermined z-dependent radius $r_m(z)$.

Verification results of this section can be found in Supplementary Demonstration of Internal Flows.

## Supplementary Experimental Method

The experiment setup is shown in Fig. S4. A reflective SLM (Holoeye GAEA-2, 3.7 um pixel pitch, 3840×2160) imprinted with computer-generated hologram patterns (the Fourier marks) transforms a collimated laser light wave at a wavelength of 532 nm into the complex field corresponding to the angular spectrum of NDPVB in the real-space coordinate system, with help of spatial filtering via a 4-F system consisting of lenses L1 and L2, and an iris as well. The resulting field is responsible for generating NDPVBs in the focal volume of lens L3 with a focal length of 200 mm. A delay line, consisting of right-angle and hollow-roof prism mirrors and a translation stage, enables the different cross-sections of NDPVB to be imaged on a complementary metal oxide semiconductor (CMOS) camera (Dhyana 400BSI, 6.5 um pixel pitch, 2048×2040) after a relay 4-F system consisting of two lenses (L4 and L5, each with a focal length of 200 mm). The combination of the delay line and the relay system enables us to record intensity cross-sections at different z-axial locations before and after the focal plane of the lens L3.

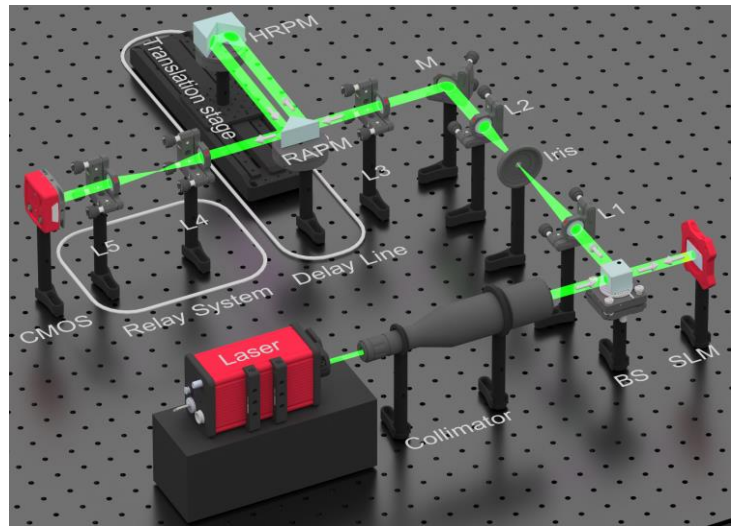

Fig. S4. Experimental setup for generating and detecting NDPVBs. BS, beam splitter; SLM, phase-only spatial light modulator; L1–L5, lens; M, mirror; RAPM, right-angle prism mirror; HRPM, hollow roof prism mirror; CMOS, complementary metal oxide semiconductor camera.

The high-order Bessel function term $J_m(\sqrt{k^2 - k_z^2} r_m)$ appears in the denominator of the synthetic angular spectrums of the customized NDPVBs like equation (10), and its zero points will make the maximum of calculated angular spectrum too large to be correctly loaded on the spatial light modulator. Our numerical simulation and optical experiment show that replacing the high-order Bessel function term with 1 can even yield fairly satisfactory customized NDPVBs. We speculate that the high-order Bessel function is merely too slightly to influence the radial amplitude profile of the ring-shaped angular spectrum and thus can be neglected with inappreciable errors.

## Supplementary Experimental Demonstration

1. **The perfection of non-diffracting NDPVB.**

Figure S5 displays the experimental results for the comparison of the traditional VBs (LGBs and HOBBs) and the NDPVBs (cf. Fig. 1a and 1b of the main text). It is clear that as topological charges increase, NDPVBs in Row 3 remain perfect, occupying the same fixed small area with a predetermined radius of 250 um. Many applications, including fiber optic data transmission, spatial OAM mode (de)multiplexing communication, and transferring larger OAM to the particles trapped in a small space in optical tweezers, will benefit from this perfection. The number of petals in the interference petal patterns between the NDPVBs and a plane wave in Row 4 justifies the topological charge carried by the NDPVBs.

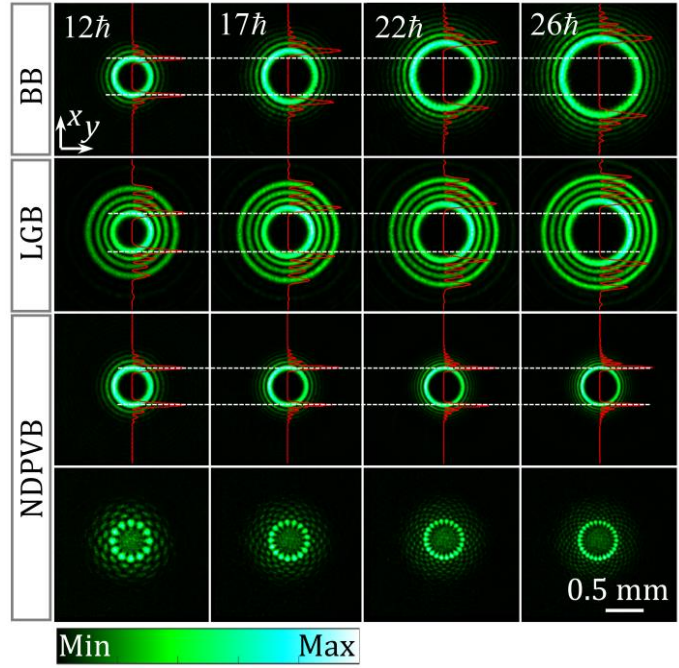

Fig. S5. Corresponding experimental results of Figs.1a-b in the main text (the red curves and the horizontal white dashed lines represent the same as Figs. 1a-b) along with the interference patterns between the NDPVBs and the plane wave in row 4.

2. **The self-healing behavior of NDPVB.**

NDPVBs exhibit self-healing behavior, in the sense that they tend to reconstitute or reform themselves even when they have been severely perturbed or impaired, as illustrated in Fig. S6 a-e, where a NDPVB encounters a square obstacle (marked by the red square) at $z = -150$ mm and self-heals when propagating. The obstacle is fabricated by photoetching a chrome square pattern on the glass substrate, forming a blocking mask. To determine the degree of reconstruction of impaired NDPVBs, we introduced the Pearson correlation coefficients (PCCs) between the intensity maps at each z-axial location after perturbation and that before perturbation at $z = -160$ mm, which is plotted in Fig. S6k. The Pearson correlation coefficients (PCCs) of matrices $X$ and $Y$ is defined by

$$PCC(X,Y) = \frac{\text{cov}(X,Y)}{\sigma_X \sigma_Y}, \tag{S16}$$

where cov(X, Y) is the covariance of $X$ and $Y$ and $\sigma_n$ ($n = X$ or $Y$) is the standard deviation of matrices $X$ or $Y$.

After the square obstacle at z = -150 mm, the PCC descends precipitously from 1.0 and rises steadily during propagation as a result of the self-healing property of the NDPVB, eventually stabilizing at roughly 0.7 under the current experimental conditions. Worth emphasizing is that the longer the non-diffracting range of the quasi-NDPVB, the more energy will be carried by the sidelobes of quasi-NDPVB and the higher value of PCC will finally stabilize at after self-healing, which is also verified by numerical tests.

For providing an insightful picture of the energy transport in this self-healing process, the momentum density has been calculated, following from the cycle-average Poynting vector[3], given as $\boldsymbol{p}=\varepsilon_0/(2\omega)\mathrm{Im}\left[\boldsymbol{U}^*\times(\nabla\times\boldsymbol{U})\right]$ where $\varepsilon_0$ and $\omega$ are the vacuum permittivity and circular frequency respectively. The transverse momentum density of Figs. S6a-e is shown in Figs. S6f-j, and the red arrows indicate the magnitude and direction of momentum density. As indicated by the evolution of the transverse momentum density, the reconstruction of the damaged intensity pattern is dominated by the inherent azimuthal momentum density and the extra radial constituent, which only emerges in the self-healing process from the outside sidelobes. The self-healing ablity of the NDPVB makes it appealing for use in disturbed turbid circumstances such as atmospheric turbulence, seawater, and biological tissues.

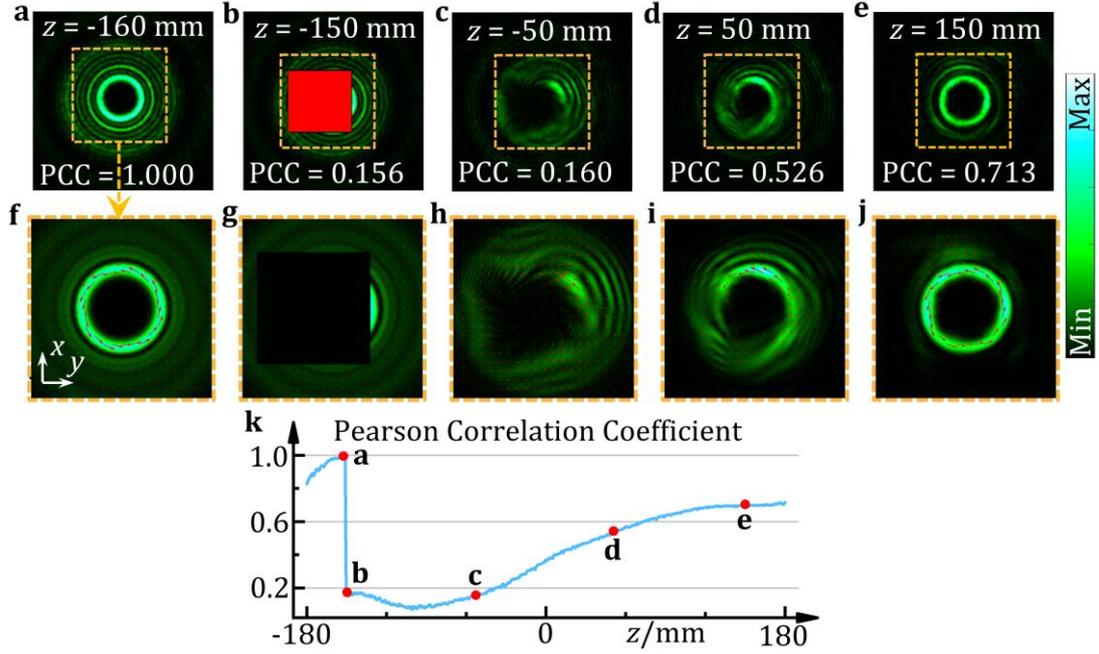

Fig. S6. Experimental demonstration of self-healing property when a NDPVB ($m$=12 and $r_0$=150$um$) is blocked by a square obstacle (red square) at $z$=-150mm: **a-e** the transversal intensity maps at different z-axial locations; **f-j** the corresponding transversal momentum density within the observed areas (orange-dashed squares) in **a-e** marked by red arrows; PCC, Pearson correlation coefficient. **k** Pearson correlation coefficient of the transversal intensity map versus z. The red dots marked in **k** represent the corresponding PCC value of **a-e**. The movies of the self-healing process are shown in the left and right panels of Movie S1 for **a-e** and **f-j**, respectively.

### 3. NDPVBs with customized propagating intensity in air and lossy media.

For verifying the validity of equation (10), we shaped three kinds of NDPVBs with $m$=12 and $r_0$=150 $um$ so that they have intensity profiles along the z-direction like the uniform $I(z)$=rect($z/2f$), linearly increased $I(z)$=0.5(-$z/f$+1), and exponentially increased $I(z)$=exp(10$z$)/exp(10$f$), respectively . The corresponding experimental results are shown in Fig. S7, which are in excellent agreement with the predesigned intensity profiles. It is seen from Figs. S7b and c (also shown in Movie S2) that the sidelobes of NDPVB function as a "reservoir" to store energy for the vortex ring and that there will be an additional radial energy flow to adjust the intensity of the vortex ring during propagation. The internal flows of customized NDPVBs is discussed in the corresponfing section of the main text.

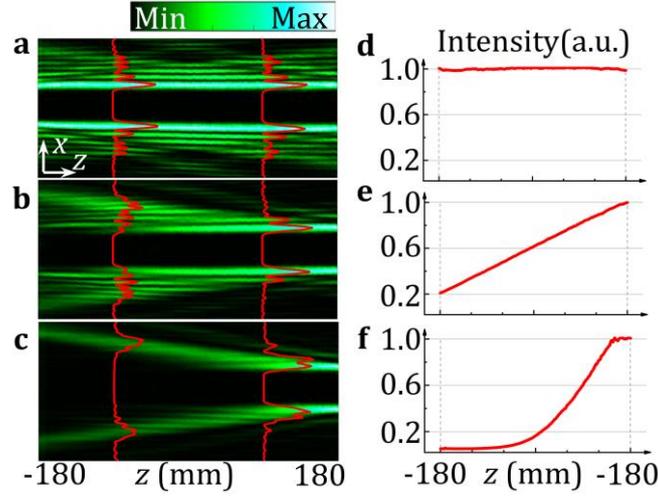

Fig. S7. Customized NDPVBs with shaped propagating intensity in the free space. *x-z* plane intensity maps of NDPVBs with predesigned intensity profiles along the z-direction: **a** uniform $I(z)=\text{rect}(z/2f)$, **b** linearly increased $I(z)=0.5(-z/f+1)$, and **c** exponentially increased $I(z)=\exp(10z)/\exp(10f)$. The red curves represent intensity profiles along the *x*-axis in the $z = -90$ mm and $z = 90$ mm planes. **d-f** the corresponding z-directional intensity profiles of the vortex rings in **a-c**. Experimental movies for *x-y* plane intensity maps in **a-c** are shown on the left, middle, and right panels of Movie S2, respectively.

Another experiment in a particular application scenario, where NDPVB propagates in lossy and scattering media like milk suspensions, has been carried out to directly elucidate the viability of shaping the propagating intensity profile. The scattered path from an incident NDPVB on a milk suspension in a glass sink is depicted in Fig. S8a and shows the attenuation of the transversal total energy during propagation. Accordingly, the scattering maps of customized NDPVBs with various propagating intensity profiles are essentially the same.

By incidenting the uncompensated NDPVB with a uniform intensity profile $I(z)=\text{rect}(z/2f)$, reflecting the transversal intensity maps at various z-axial locations by a mirror inserted in the sink, and recording them by a COMS, we were able to determine the decay curve of this milk suspension. The resulting transversal intensity maps in the lossy medium are shown in Fig. S8b and the measured decay curve roughly equates to $\exp(-3.8z)$ (the blue-dashed curve in Fig. S8d). Similar to this, the transversal intensity maps at different z-axial locations of the incident attenuation-compensated NDPVB with the exponentially increased intensity profile $I(z)=\exp(3.8z)/\exp(3.8f)$ in this lossy medium are captured and shown in Fig. S8c. Values of the normalized intensity on the vortex ring of the uncompensated and attenuation-compensated NDPVBs at different z-axial locations have been marked on Fig. S8b and c with red font and plotted in Fig. S8d with the blue triangles and blue circles respectively. The sidelobes of the attenuation-compensated NDPVB in Fig. S8c compensate for the lossy energy on the vortex ring when propagating in the attenuating medium, and as a result, the intensity profile on the vortex ring along the z-direction can remain almost uniform, despite the fact that the transversal total energy is also attenuating.

Another milk suspension with a different concentration is substituted into the sink to demonstrate the viability of attenuation compensation in various media, and the decay curve of the milk suspension as measured by uncompensated NDPVB roughly equals $\exp(-5.1z)$ (the red-dashed curve in Fig. S8d). Plotted in Fig. S8d with red triangles and red circles, respectively, are the intensity profiles on the vortex rings of the uncompensated NDPVB and the attenuation-compensated NDPVB with a predesigned profile $I(z)=\exp(5.1z)/\exp(5.1f)$ in this milk suspension; the latter remains almost constant as expected as well. The scattered paths and the transversal intensity maps of the no-compensated and attenuation-compensated NDPVBs in a milk suspension with different concentrations are similar to those in Fig. S8a-c and are therefore omitted here. The attenuation-compensated NDPVBs remaining the constant propagating intensity in different media will benefit a variety of optical applications under disturbed turbid circumstances, e.g. underwater communication, manipulation in some solutions, and imaging for biological samples.

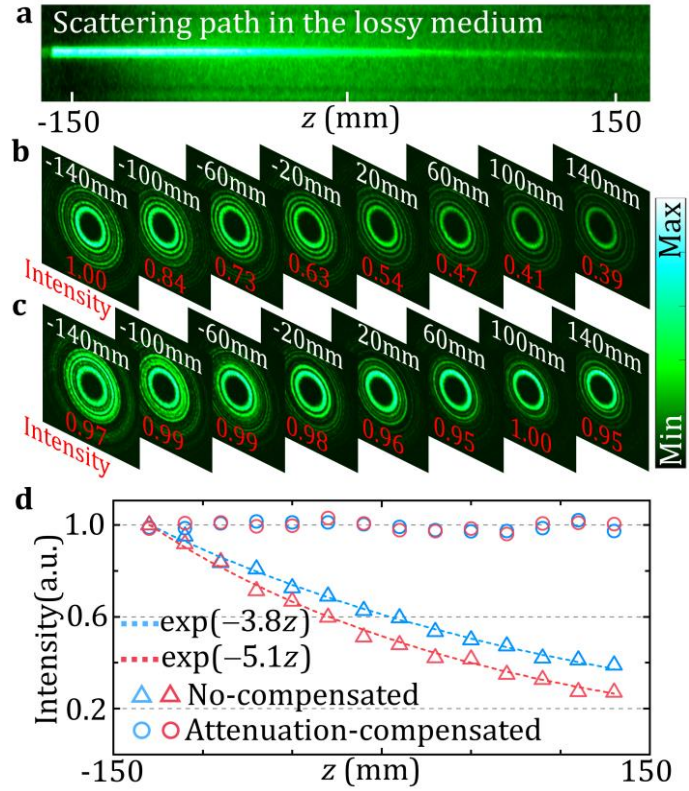

Fig. S8. Attenuation-compensated NDPVBs in lossy media. The scattered path when a NDPVB propagates attenuated in the lossy medium (milk suspension with measured attenuation curve exp(-3.8$z$) plotted in **d** with the blue-dashed curve). x-y plane intensity maps of **a** the no-compensated NDPVB with a predesigned uniform propagating intensity $I(z)=\text{rect}(z/2f)$ (e.g. Fig. 1**c**) and **b** the attenuation-compensated NDPVB with a predesigned exponentially increased propagating intensity $I(z)=\exp(3.8z)/\exp(3.8f)$ in this lossy medium at $z = -140$ mm,-100 mm,-60 mm, -20 mm, 20 mm, 60 mm, 100 mm, and 140 mm respectively. The red font digits on the maps depict the normalized intensity value on the vortex rings of the no-compensated and attenuation-compensated NDPVBs which are plotted in **d** with the blue triangles and blue circles, respectively. The red triangles and red circles in **d** represent the measured longitudinal intensity profiles of the no-compensated $I(z)=\text{rect}(z/2f)$ and attenuation-compensated $I(z)=\exp(5.1z)/\exp(5.1f)$ NDPVBs, respectively, in another milk suspension with measured attenuation curve exp(-5.1$z$) (the red-dashed curve).

### 4. The self-accelerating and self-similar behavior of customized NDPVB.

Self-accelerating NDPVB can act as a snowblower conveying particles from one compartment to another[4] or convey information to the obscured invisible receiver[5]. A rectilinearly propagating NDPVB ($m$=12, $r_0$=150$um$, and $I(z)=\text{rect}(z/2f)$) obscured by a "wall" at $z$=0 plane is severely damaged shown in Fig. S9a whereas the self-accelerating NDPVB with self-bending trajectory $s(z) = 1.8\times10^{-4}\,[1-z^2/f^2,\,0]$ generated from equation (12) can surmount this wall illustrated in Fig. S9b with maintaining the transversal pattern and predesigned propagating intensity. When a "receiver" is completely obscured by a "plate" on the optical axis, the NDPVB (having $m$=12, $r_0$=100$um$, and $I(z)$=rect($z/2f$)) cannot be received, as shown in Fig. S9c, but the self-similar NDPVB with the tailored radius $r_m(z)=1\times10^{-4}-1.7\times10^{-4}\,(z^2/f^2-1)$, which is generated by equation (16), can easily bypasd this "plate" and arrive the obscured "receiver", as shown in Fig. S9d. These self-accelerating and self-similar NDPVBs with controllable propagating intensity and phase at will expands the "toolkit" of beams' shaping and will be utilized in a variety of applications, including filamentation, particle manipulation, biomedical imaging, plasmons, and material processing, among others.

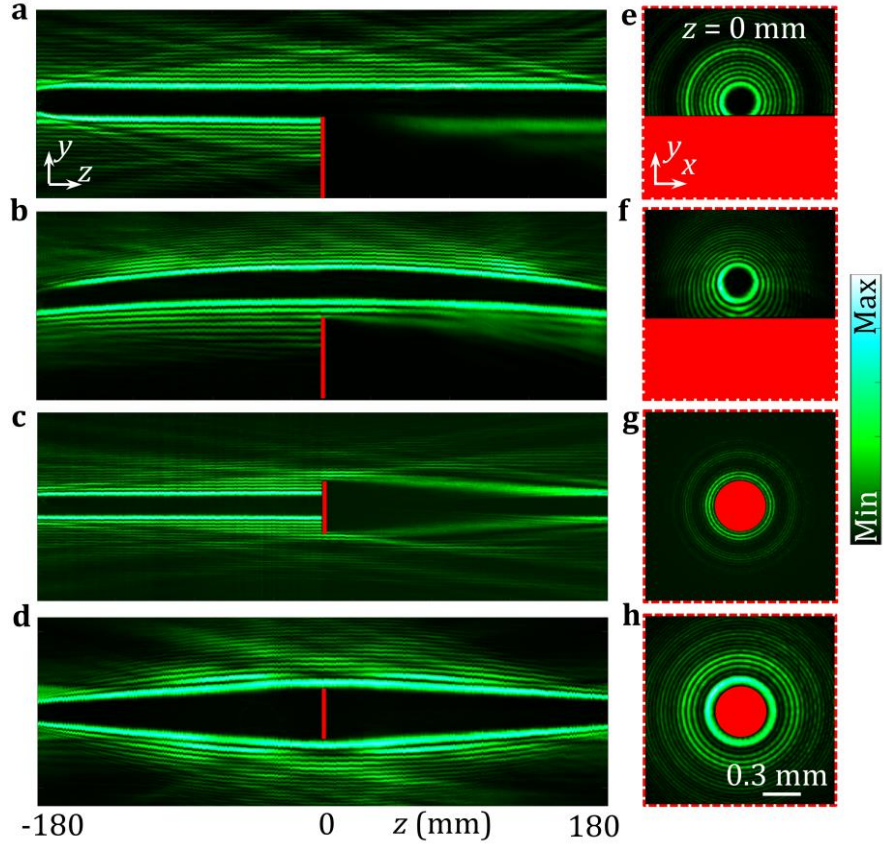

Fig. S9. Experimental demonstrations for the self-accelerating and self-similar behaviors of customized NDPVBs: **a** Intensity maps in the *y-z* plane of a rectilinearly propagating NDPVB obstructed by a 'wall' (red-solid line) at *z* = 0mm; **b** the self-accelerating NDPVB self-bends to surmount this 'wall' (Movie S3); **c** Intensity maps in the *y-z* plane of NDPVB obstructed by a 'plate' (red-solid line) at *z*=0mm; **d** the self-similar NDPVB self-stretches to "swallow" this 'plate' (Movie S4). **e-h** Intensity maps in the x-y plane at the obstacle plane *z*=0mm (the position of the red-solid lines marked in **a-d**). The sharp-edged red rectangles and plates in **e-h** depict the obstacles, the "wall" and "plate", which are experimentally realized by masks with photoetching chrome patterns on the glass substrate.

## 5. The self-rotating behavior and optical ring lattice of structured NDPVBs.

Experimental results of self-rotating NDABLBs and NDADLBs are demonstrated in Fig. S10 and are in excellent agreement with the theoretical predictions. The adjustable intensity gradient of the NDRBLB in Fig. S10b is much larger than that in Fig. S10a because of the larger *N* with the same radius $r_0$.

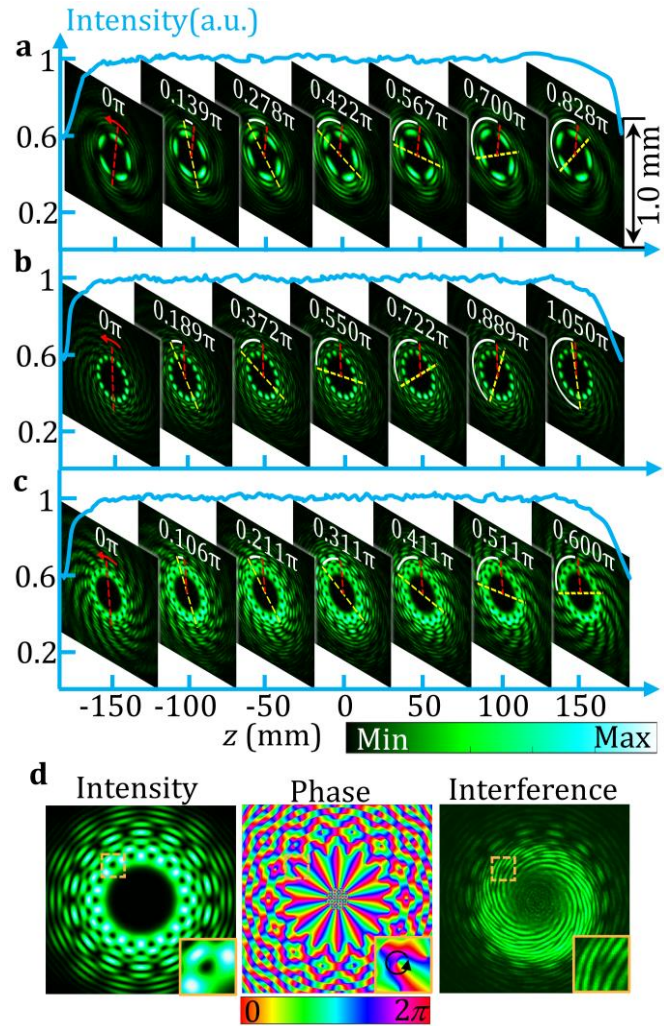

Fig. S10. The self-rotating propagation of the non-diffracting annular bright lattice beam (NDABLB) with **a** ($r_0$ = 411.4 um, $m$ = 12, $n$ = 17) resulting in $N$ = 5 and $\omega(z)$ = -8.54 rad/m, **b** ($r_0$ = 411.4 um, $m$ = 12, $n$ = 26) resulting in $N$ = 14 and $\omega(z)$ = -10.98 rad/m, and **c** the non-diffracting annular dark lattice beam (NDADLB) with ($r_m$ = 411.4 um, $r_n$ = 503.2 um, $m$ = 12, $n$ = 26) resulting in $N$ = 14 and $\omega(z)$ = -6.21 rad/m, respectively. The blue-solid curves represent the z-directional intensity profiles; the red arrows depict the direction of rotation; the red-dashed lines serve as a reference of the orientation at z = -150 mm; the yellow-dashed lines depict the orientations after self-rotating and the white-solid arcs mark the variations of orientation from z = 150 mm. **d** the x-y plane distribution of the NDADLB at z = 0 in **c**: left, the intensity map; middle, the phase map; right, the experimental interference pattern with a spherical reference wave. The insets in **d** are enlarged views of the orange-dashed areas. Experimental movies for **a-c** are shown on the left, middle, and right panels of Movie S5, respectively.

## Supplementary Demonstration of Internal Flows

When only shaping the propagating intensity $I(z)$ with $g(z) = h(z) = 0$ and $r_m(z) = r_0$, the equation (31) can be simplified as

$$\boldsymbol{p}_\perp(r_0,\varphi,z)=I(z)\frac{m}{r_0}\hat{\boldsymbol{\varphi}}. \tag{S17}$$

The ideal NDPVB depicted by equation (4), which is always propagation-invariant, will not exhibit the radial momentum density. For physically feasible quasi-NDPVB truncated in a limited z-range, however, there inevitably exhibits an extra radial momentum density, whcih dominates the formation, the propagating intensity evolution, and the dissipation of the quasi-NDPVB. According to the continuity equation[3], the transverse momentum density impacts how the intensity structure of beams propagates, and the relationship can be expressed as

$$\frac{\partial w(r,\varphi,z)}{\partial z} = -c(\nabla_\perp \boldsymbol{p}_\perp(r,\varphi,z)), \tag{S18}$$

where $w(r, \varphi, z)$ is the energy density and $c$ is the velocity of light. For the vortex ring at $r = r_0$, $\partial w(r_0, \varphi, z) / \partial z$ is proportional to $I'(z)$ and the divergence of transverse momentum density $\nabla_\perp \boldsymbol{p}_\perp(r_0,\varphi,z)$ can be thought of as the flux of transverse momentum density acrross the vortex ring. Since the inherent azimuthal momentum density is always circulating inside the ring and thus will not introduce flux, the flux across the vortex ring $\nabla_\perp \boldsymbol{p}_\perp(r_0,\varphi,z)$ is entirely contributed by the radial momentum density $p_r$ (i.e. proportional to $p_r$) because of the axisymmetry. Hence, we can derive $p_r = -C_5 I'(z)$, where $C_5$ is a positive constant, from equation (S18), and the transversal momentum density on the vortex ring of NDPVBs with the customized propagating intensity $I(z)$ is given by

$$\boldsymbol{p}_\perp(r_0,\varphi,z)=I(z)\frac{m}{r_0}\hat{\boldsymbol{\varphi}}(z) - C_5 I'(z)\hat{\boldsymbol{r}} = p_\varphi \hat{\boldsymbol{\varphi}} + p_r \hat{\boldsymbol{r}}. \tag{S19}$$

For instance, a negative momentum density tends to propel the energy flowing inward to the vortex ring from outside sidelobes for $I'(z)>0$), and vice versa for $I'(z)<0$. Theoretical predictions from equation (S19) have been verified in Fig. S11, which are well consistent with the numerical outcomes from $\boldsymbol{p} = \varepsilon_0/(2\omega)\mathrm{Im}\left[\boldsymbol{U}^* \times (\nabla \times \boldsymbol{U})\right]$.

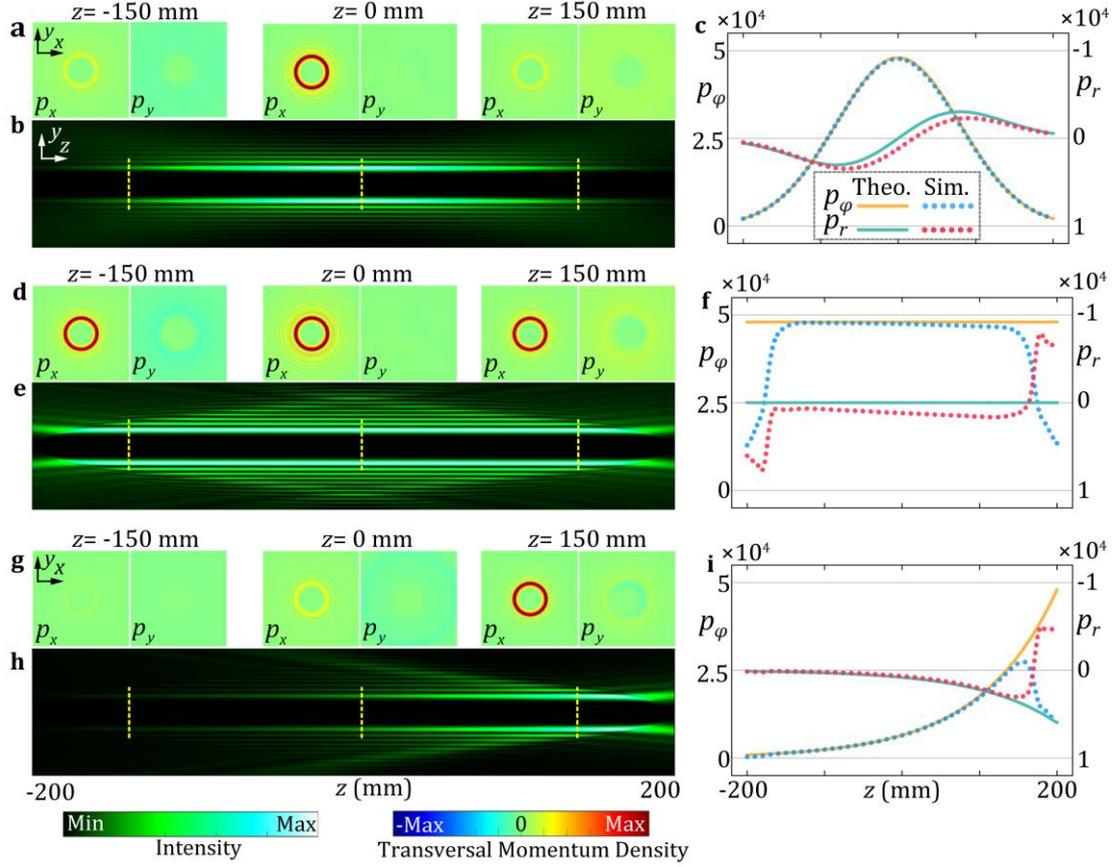

Fig. S11. Evolving transversal momentum density of NDPVBs ($m = 12$ and $r_0 = 150$ um) with the customized propagating intensity $I(z)$. **a** the transversal momentum density of the customized NDPVB with Gaussian intensity profile $I(z) = \exp(-z^2/0.0128)$ in $x$-$y$ planes at $z = -150$ mm, 0 mm, 150 mm whose locations are marked by yellow-dashed lines in the intensity map **b** in the $y$-$z$ plane. The evolving radial ($p_r$) and azimuthal ($p_\varphi$) constituents of the transversal momentum density on the vortex ring are plotted in **c** where the solid curves depict the theoretical predictions from equation (S19), the dotted curves depict the numerical results from $\bm{p} = \varepsilon_0/(2\omega)\text{Im}\left[\bm{U}^* \times (\nabla \times \bm{U})\right]$ while the maximum intensity of NDPVB is normalized. **d-f** and **g-i** are the same as in **a-c** but for uniform $I(z)=\text{rect}(z/2f)$ and exponentially increasing $I(z)=\exp(10z)/\exp(10f)$ intensity profiles. The movies of the evolving transversal momentum densities for (**a-c**), (**d-f**), and (**g-i**) are shown in Movie S7-S9, respectively.

When only shaping the self-similar behavior with changing radius $r_m(z)$ while $I(z)=1$ and $g(z)=h(z)=0$, the equation (31) can be simplified as

$$\bm{p}_\perp(r_m(z),\varphi,z) = \frac{m}{r_m(z)}\hat{\bm{\varphi}} + k_0 r_m'(z)\hat{\bm{r}} = p_\varphi \hat{\bm{\varphi}} + p_r \hat{\bm{r}}. \tag{S20}$$

Theoretical predictions from equation (S20) have been verified in Fig. S12, which are well consistent with the numerical outcomes from $\bm{p} = \varepsilon_0/(2\omega)\text{Im}\left[\bm{U}^* \times (\nabla \times \bm{U})\right]$.

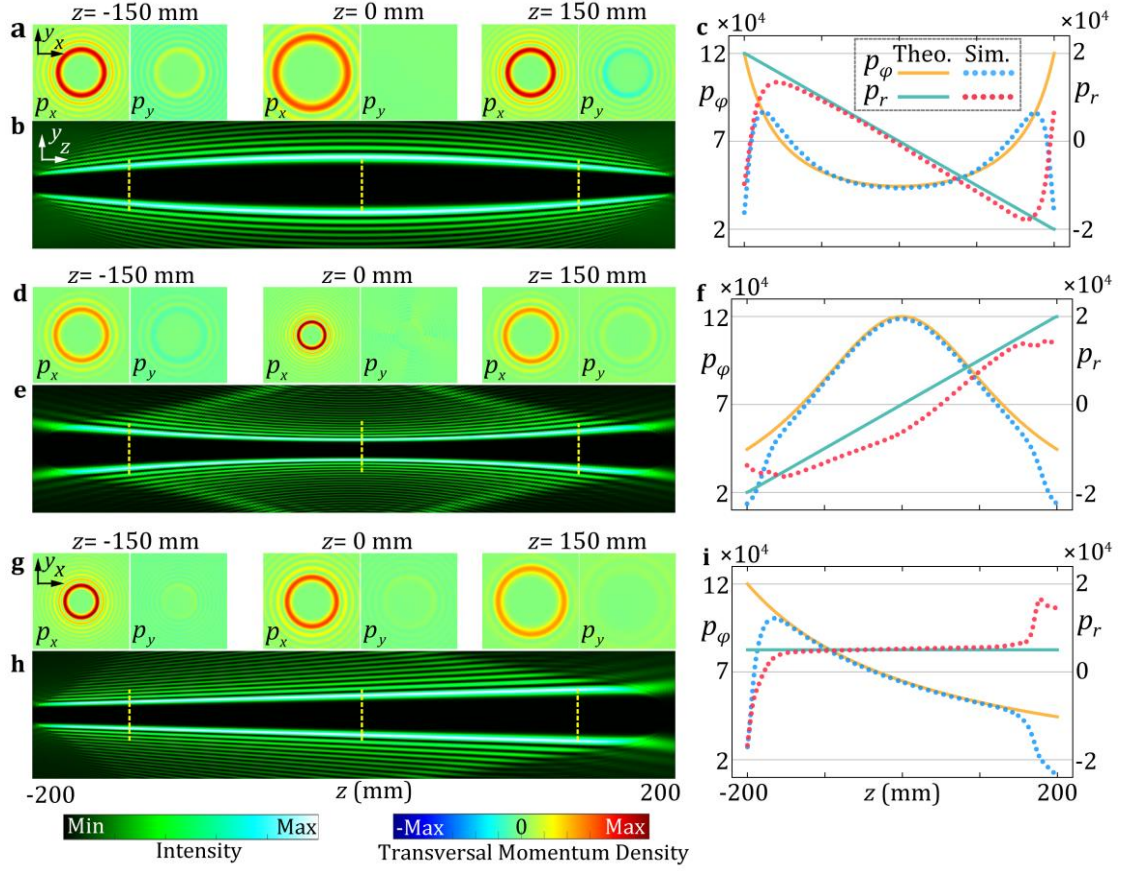

Fig. S12. Evolving transversal momentum density of NDPVBs ($m$=12) with customized radius $r_m(z)$. **a** the transversal momentum density of the customized NDPVB with self-stretching radius $r_m(z)$=100e$^{-6}$-170e$^{-6}$($z^2/f^2$-1) in $x$-$y$ planes at $z$ = -150 mm, 0 mm, 150 mm whose locations are marked by yellow-dashed lines in the intensity map **b** in the $y$-$z$ plane. The evolving radial ($p_r$) and azimuthal ($p_\varphi$) constituents of the transversal momentum density on the vortex ring are plotted in **c** where the solid curves depict the theoretical predictions from equation S20, the dotted curves depict the numerical results from $\boldsymbol{p}=\varepsilon_0/(2\omega)\text{Im}\left[\boldsymbol{U}^*\times(\nabla\times\boldsymbol{U})\right]$ while the maximum intensity of NDPVB is normalized. **d-f** and **g-i** are the same as in **a-c** but for self-shrinking radius $r_m(z)$ = 270e$^{-6}$+170e$^{-6}$($z^2/f^2$-1) and linearly stretching radius $r_m(z)$ = 100e$^{-6}$+85e$^{-6}$($z/f$+1). The movies of the evolving transversal momentum densities for **(a-c)**, **(d-f)**, and **(g-i)** are shown in Movie S10-S12 respectively.

When only shaping the self-accelerating behavior with trajectory $\boldsymbol{s}(z)=(g(z), h(z))$ while $I(z)=1$ and $r_m(z)=r_0$, the equation (31) can be simplified as

$$\boldsymbol{p}_\perp(r_0,\varphi,z)=\frac{m}{r_0}\hat{\boldsymbol{\varphi}}+k_0 g'(z)\hat{\boldsymbol{x}}+k_0 h'(z)\hat{\boldsymbol{y}}. \tag{S21}$$

In contrast to the above two examples with axisymmetry, the transversal momentum density on the vortex ring of self-accelerating NDPVBs is non-axisymmetric and should be described in Cartesian coordinates. Equation (S21) can be rewritten as

$$\boldsymbol{p}_\perp(r_0,\varphi,z)=(k_0 g'(z)-\frac{m}{r_0}\sin\theta)\hat{\boldsymbol{x}}+(k_0 h'(z)+\frac{m}{r_0}\cos\theta)\hat{\boldsymbol{y}}=(p_x,p_y), \tag{S22}$$

which shows that the $x$- and $y$- components of the momentum density $(m/r_0)\sin\theta$, $(m/r_0)\cos\theta$ varying with $\theta$ on the vortex ring have an impact on the original momentum density $(k_0g'(z), k_0h'(z))$ due to the beams' self-accelerating evolution. For unambiguous demonstration, we used the 0th-order Bessel beam ($m=0$), which has no azimuthal momentum density, to validate the momentum density induced by self-accelerating behavior, the results are shown in Fig. S13, which are in good agreement with the numerical results from $\boldsymbol{p}=\varepsilon_0/(2\omega)\mathrm{Im}\left[\boldsymbol{U}^*\times(\nabla\times\boldsymbol{U})\right]$.

We also confirmed the equation (S22) with $m\neq0$: the maximum and minimum of $p_x$ on the vortex ring are $k_0g'(z)+(m/r_0)$ and $k_0g'(z)-(m/r_0)$ while the maximum and minimum of $p_y$ on the vortex ring are $k_0h'(z)+(m/r_0)$ and $k_0h'(z)-(m/r_0)$. Adding the maximum and minimum values of $p_x$ and $p_y$ yields $2k_0g'(z)$ and $2k_0h'(z)$, respectively, which well match the numerical results as well.

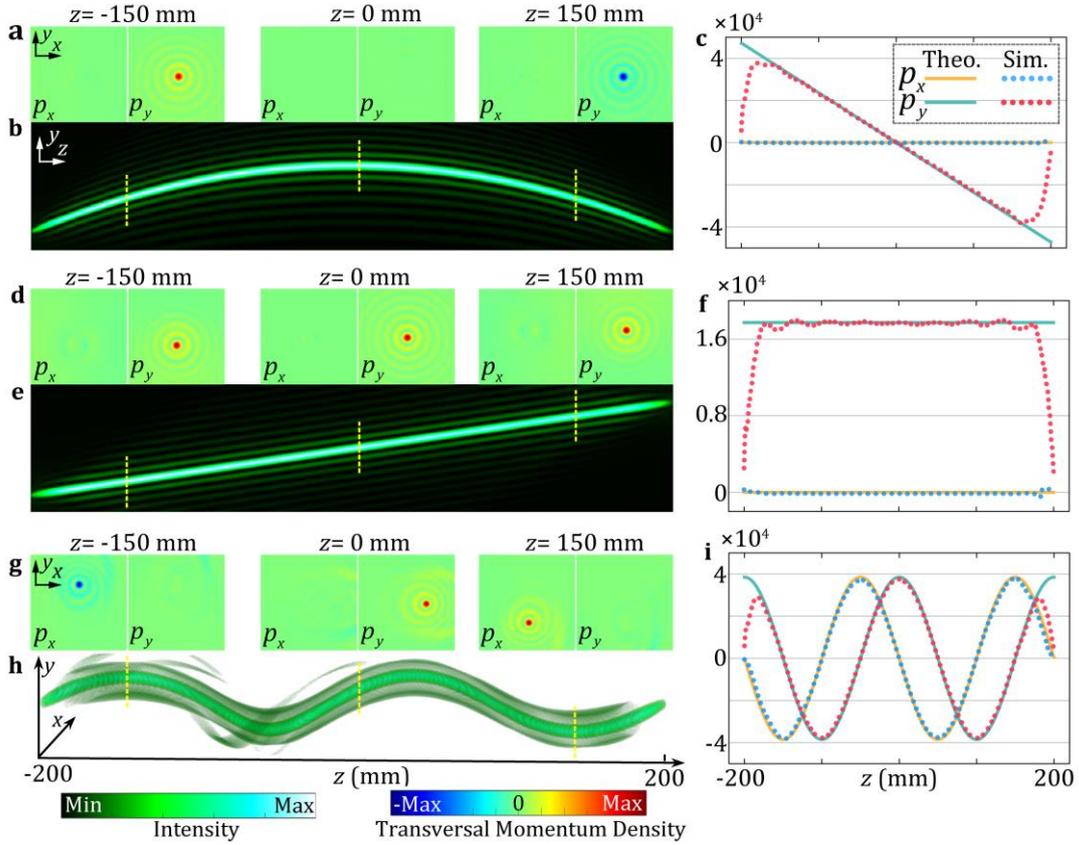

Fig. S13. Evolving transversal momentum density of 0th-order Bessel beam with self-accelerating trajectory $\boldsymbol{s}(z)=(g(z), h(z))$. **a** the transversal momentum density of the customized Bessel beam with parabolic trajectory $\boldsymbol{s}(z)= 4\times10^{-4} (0, 1-z^2/f^2)$ in $x$-$y$ planes at $z = $ -150 mm, 0 mm, 150 mm whose locations are marked by yellow-dashed lines in the intensity map **b** in the $y$-$z$ plane. The evolving $x$ ($p_x$) and $y$ ($p_y$) constituents of the transversal momentum density on the mainlobe are plotted in **c**, where the solid curves depict the theoretical predictions from equation S21 and the dotted curves depict the numerical results from $\boldsymbol{p}=\varepsilon_0/(2\omega)\mathrm{Im}\left[\boldsymbol{U}^*\times(\nabla\times\boldsymbol{U})\right]$ while the maximum intensity of the 0th-order Bessel beam is normalized. **d-f** and **g-i** are the same as in **a-c** but for linear $\boldsymbol{s}(z)= 3\times10^{-4} (0, z/f)$ and spiral $\boldsymbol{s}(z)= 1.038\times10^{-4} [\cos(2\pi z/f), \sin(2\pi z/f)]$ trajectories. The movies of the evolving transversal momentum densities for (**a-c**), (**d-f**), and (**g-i**) are shown in Movie S13-S15, respectively.

## Supplementary Movies

Movie S1. Experimental movies for the self-healing evolution of NDPVB: (left) transversal intensity maps in Figs. S6a-e, (right) transversal energy flow densities in Figs. S6f-j.

Movie S2. Experimental movies for $x$-$y$ plane intensity maps of NDPVBs with predesigned intensity profiles along the z-direction intensity profiles in Figs. S7a-c: (left) uniform $I(z)$ = rect$(z/2f)$, (middle) linearly increased $I(z) = 0.5(-z/f+1)$, and (right) exponentially increased $I(z) = \exp(10z)/\exp(10f)$.

Movie S3. Experimental movies for NDPVBs with predesigned trajectories and obstructed by a 'wall' at $z = 0$ mm: (left) the rectilinearly propagating NDPVB in Fig. S9a with ($m = 12$, $r_0 = 150$ um, $I(z)$ = rect$(z/2f)$), (right) the self-accelerating NDPVB in Fig. S9a with $s(z) = 1.8 \times 10^{-4} [1-z^2/f^2, 0]$ that self-bends to surmount this 'wall'.

Movie S4. Experimental movies for the self-similar NDPVBs obstructed by a 'plate' at $z = 0$mm: (left) the NDPVB in Fig. S9c with ($m = 12$, $r_0 = 100$ um, $I(z)$ = rect$(z/2f)$), (right) the self-similar NDPVB in Fig. S9d with $r_m(z) = 1 \times 10^{-4} - 1.7 \times 10^{-4} (z^2/f^2 - 1)$ that self-stretches to "swallow" this 'plate'.

Movie S5. Experimental movies for the non-diffracting annular lattice beams: (left) the NDRBLB in Fig. S10a ($r_0 = 411.4$ um, $m = 12$, $n = 17$) resulting in $N = 5$ and $\omega(z) = -8.54$ rad/m, (middle) the NDRBLB in Fig. S10b ($r_0 = 411.4$ um, $m = 12$, $n = 26$) resulting in $N = 14$ and $\omega(z) = -10.98$ rad/m, (right) the NDRDLB in Fig. S10c ($r_m = 411.4$ um, $r_n = 503.2$ um, $m = 12$, $n = 26$) resulting in $N = 14$ and $\omega(z) = -6.21$ rad/m.

Movie S6. The complex distribution of the TC-customized NDPVB with $m(z) = -12\text{rect}(z/f+0.5)+17\text{rect}(z/f-0.5)$ (left) and the NDABLB resulting from the coaxial interference between it and the NDPVB with $m = 8$ (right).

Movie S7. Evolving radial and azimuthal momentum densities ($p_r$ and $p_\varphi$) of the NDPVB ($m=12$ and $r_0=150$um) shown in Figs. S11a-c with a Gaussian intensity profile of $I(z)=\exp(-z^2/0.0128)$.

Movie S8. Evolving radial and azimuthal momentum densities ($p_r$ and $p_\varphi$) of the customized NDPVB ($m = 12$ and $r_0 = 150$ um) shown in Figs. S11d-f with a uniform intensity profile of $I(z)$ = rect$(z/2f)$.

Movie S9. Evolving radial and azimuthal momentum densities ($p_r$ and $p_\varphi$) of the customized NDPVB ($m = 12$ and $r_0 = 150$ um) shown in Figs. S11g-i with an exponentially increasing intensity profile of $I(z) = \exp(10z)/\exp(10f)$.

Movie S10. Evolving radial and azimuthal momentum densities ($p_r$ and $p_\varphi$) of the customized NDPVB ($m = 12$) shown in Figs. S12a-c with a self-stretching radius of $r_m(z) = 100e^{-6}-170e^{-6}(z^2/f^2-1)$.

Movie S11. Evolving radial and azimuthal momentum densities ($p_r$ and $p_\varphi$) of the customized NDPVB ($m = 12$) shown in Figs. S12d-f with a self-shrinking radius of $r_m(z) = 270e^{-6}+170e^{-6}(z^2/f^2-1)$.

Movie S12. Evolving radial and azimuthal momentum densities ($p_r$ and $p_\varphi$) of the customized NDPVB ($m = 12$) shown in Figs. S12g-i with a linearly stretching radius of $r_m(z) = 100e^{-6}+85e^{-6}(z/f+1)$.

Movie S13. Evolving $x$- and $y$-directional momentum densities ($p_x$ and $p_y$) of the 0th Bessel beam shown in Figs. S13a-c with a parabolic trajectory of $s(z) = 4 \times 10^{-4} (0, 1-z^2/f^2)$.

Movie S14. Evolving $x$- and $y$-directional momentum densities ($p_x$ and $p_y$) of the 0th Bessel beam shown in Figs. S13d-f with a linear trajectory of $s(z) = 3 \times 10^{-4} (0, z/f)$.

Movie S15. Evolving $x$- and $y$-directional momentum densities ($p_x$ and $p_y$) of the 0th Bessel beam shown in Fig. S13g-i with a spiral trajectory of $s(z) = 1.038 \times 10^{-4} [\cos(2\pi z/f), \sin(2\pi z/f)]$.

## Supplementary Reference